\documentclass[iop]{emulateapj}

\usepackage{footmisc}
\usepackage{amsmath}
\usepackage{graphicx,epsfig,epsf,graphics,epstopdf}
\usepackage{natbib,hyperref}
\def\apj{ApJ}
\def\nat{Nature}
\def\mnras{MNRAS}
\def\apjl{ApJL}
\def\apjs{ApJS}
\def\araa{ARAA}
\def\aap{A\&A}
\def\aj{AJ}

\shorttitle{Disequilibrium Chemistry with JWST}
\shortauthors{Blumenthal et al.}

\begin{document}

\title{A Comparison of Simulated JWST Observations Derived from Equilibrium and Non-Equilibrium Chemistry Models of Giant Exoplanets}

\author{Sarah D. Blumenthal\altaffilmark{1, 2, 3}, Avi M. Mandell\altaffilmark{1}, Eric H\'ebrard\altaffilmark{1, 3}, Natasha E. Batalha\altaffilmark{1, 4}, Patricio E. Cubillos\altaffilmark{5}, Sarah Rugheimer\altaffilmark{6}, Hannah R. Wakeford\altaffilmark{1}}

\submitted{Accepted to ApJ November 27, 2017}

\altaffiltext{1}{NASA Goddard Space Flight Center, Center for Astrobiology, Greenbelt, MD 20771}
\altaffiltext{2}{University of Maryland Baltimore Country, CRESST}
\altaffiltext{3}{Astrophysics Group, School of Physics, University of Exeter, Stocker Road, Exeter EX4 4QL, UK}
\altaffiltext{4}{Department of Astronomy \& Astrophysics, Pennsylvania State University, State College, PA 16801}
\altaffiltext{5}{Space Research Institute, Austrian Academy of Sciences, Schmiedlstra{\ss}e 6, 8042 Graz, Austria}
\altaffiltext{6}{School of Geography \& Geosciences, University of St Andrews, St Andrews, United Kingdom}

\begin{abstract} 
We aim to see if the difference between equilibrium and disequilibrium chemistry is observable in the atmospheres of transiting planets by the James Webb Space Telescope (JWST). We perform a case study comparing the dayside emission spectra of three planets like HD 189733b, WASP-80b, and GJ436b, in and out of chemical equilibrium at two metallicities each. These three planets were chosen because they span a large range of planetary masses and equilibrium temperatures, from hot and Jupiter-sized to warm and Neptune-sized. We link the one-dimensional disequilibrium chemistry model from \citet{Venot2012} in which thermochemical kinetics, vertical transport, and photochemistry are taken into account, to the one-dimensional, pseudo line-by-line radiative transfer model, \texttt{Pyrat Bay}, developed especially for hot Jupiters, and then simulate JWST spectra using \texttt{PandExo} for comparing the effects of temperature, metallicity, and radius. We find the most significant differences from 4 to 5 $\mu$m due to disequilibrium from CO and CO$_{2}$ abundances, and also H$_{2}$O for select cases. Our case study shows a certain ``sweet spot'' of planetary mass, temperature, and metallicity where the difference between equilibrium and disequilibrium is observable. For a planet similar to WASP-80b, JWST's NIRSpec G395M can detect differences due to disequilibrium chemistry with one eclipse event. For a planet similar to GJ 436b, the observability of differences due to disequilibrium chemistry is possible at low metallicity given five eclipse events, but not possible at the higher metallicity. 
 
\end{abstract}

\section{Introduction}
\label{sec:intro}

The total exoplanet count currently exceeds 3,000, and more than 700 orbit stars with a J magnitude between 6 and 12 (http://exoplanets.org) This subset represents optimal targets for observation with \textit{The James Webb Space Telescope} (JWST)  because their stellar fluxes do not saturate the detectors in standard time-series modes, and are therefore prime candidates for exploring composition and structure of exoplanet atmospheres. Now, we are entering the era of JWST, which will revolutionize the characterization of exoplanet atmospheres. Launching October of 2018, JWST is equipped with a 6.5 m primary mirror and large wavelength coverage from 0.6 to 28 $\mu$m with 3 near-infrared instruments and 1 mid-infrared instrument. Each observing mode offers higher spectral resolving power (R=100-2700) as compared to current space-based facilities (R$<$100).
 
The optimistic lifetime of JWST is 10 years, determined by its supply of on-board fuel. The current state-of-the-art space telescope, the \textit{Hubble Space Telescope} (HST), is estimated to be decommissioned some time in 2020, making the operational lifetime of this serviceable telescope 30 years (since it's launch in 1990). Due to the limited lifetime of JWST in contrast to HST, it is important the community exploits the full potential of JWST as early as possible to produce a high scientific yield to expand our current understanding, to strongly motivate the commissioning of future space-based telescopes, and provide clear improvements for the continued creation of these type of missions, such as the \textit{Wide-Field Infrared Survey Telescope} (WFIRST) and the \textit{Large UV/Optical/Infrared Surveyor} (LUVOIR). However, determining the best strategy for utilizing time with JWST is unclear, as mentioned in \citet{Stevenson2016}. \citet{Stevenson2016} identified WASP-62b as a promising target for Early Release Science with JWST. This target has a high signal-to-noise ratio and an effective temperature $\sim$1400 K. As this is a relatively hot, it would be expected that the chemistry of this planet would be dominated by thermochemical equilibrium. For lower temperatures, however, it would be expected that the chemistry of a planet would be out equilibrium. This is seen in our own solar system. 
 
Much of our early understanding of the chemical composition of the atmospheres of highly irradiated exoplanets was gleaned using the assumption of chemical equilibrium.  From the precedent set by earlier work \citep{Burrows1997, BurrowsnSharp1999, SeagernSasselov2000, LoddersnFegley2002, Burrows2008, Showman2009}, equilibrium chemistry has become a widely used module in the analysis of exoplanet atmospheres. Most thermochemical equilibrium models employ a Gibbs' free energy minimization strategy of iterating over adjustments of chemical abundance using thermodynamic data to achieve convergence. Studies have frequently concluded that disequilibrium processes could aid in a better interpretation of observational data, for example \citet{Fortney2006, Showman2009, Fortney2010}. This motivated development of more sophisticated chemistry models that consider reaction timescales--- chemical kinetics, and photochemistry \citep{Liang2003, CooperShowman2006, Zahnle2009, Line2010, Moses2011, Venot2012, Venot2016}. Additionally, \citet{Baraffe2014} discusses the brown dwarf, cousin to giant exoplanets, and points out the importance of non-equilibrium chemistry in brown dwarfs. 

In this study, we focus on the impact of disequilibrium chemistry on the abundances of molecular species in the atmospheres of planets that may be observable by JWST. We choose to base our planetary parameters on several known transiting planets likely to be observed with JWST, spanning a range in planetary size and effective temperature (effectively signal-to-noise ratio) from hot Jupiter to warm sub-Neptune. We develop models based on the bulk properties of several well-known transiting planets, HD 189733b, WASP-80b, and GJ 436b. We explore the effect of metallicity assuming two different metallicities. The well-known planets HD 189733b and GJ 436b are chosen as representative bookends of our study. We chose the intermediate target WASP-80b because it has been described as the ``missing link'' in \citet{Triaud2015}, as it receives the same flux as GJ 436b at 3.6 $\mu$m \citep{Triaud2015} but has a radius similar to Jupiter. Additionally, because WASP-80b has moderate temperatures ($\sim$800 K) and its host star exhibits strong chromospheric activity, it is an excellent target for studying disequilibrium chemistry effects. Thus, the goal of this study to begin to define a regime where distinguishing disequilibrium chemistry from equilibrium chemistry may be possible.  

\section{Method}
\label{sec:method}

For the first time we link together the equilibrium and disequilibrium chemistry models of \citet{Venot2012} to the open-source radiative transfer code \texttt{Pyrat Bay} and then to the open-source JWST simulator \texttt{PandExo} to simulate 1D globally-averaged secondary eclipse spectra at JWST resolution. It is the first study of its kind to compare disequilibrium and equilibrium chemistry on synthetic JWST spectra.

We chose to conduct our study in 1D rather than in 3D in order to explore a broader parameter space of different radii, temperatures, and metallicities using a robust chemical scheme. Exploring this same parameter space in 3D would be very computationally expensive to employ such a detailed chemical scheme. However, we recognize that 1D has its limitations as a proxy for 3D. \citet{Fortney2010} shows that for HD 189733b transmission spectra, 1D is a good proxy for 3D, but for HD 209458b transmission spectra, 1D is not a good a proxy due to the temperature contrast generated by the terminator region. 

For our study, our overall modelling scheme is as follows:
\begin{enumerate}
    \item For HD 189733b, we use the temperature-pressure profile from \citet{Moses2011}. For WASP-80b, and GJ 436b,  we generate a temperature-pressure profile using \citet{Parmentiermodel} and heat the upper atmosphere according to the process outlined in \citet{Moses2011}.
    \item We solve for the altitude of each temperature-pressure point from our profile assuming hydrostatic equilibrium. 
    \item We run the equilibrium chemistry model using the inputs of the calculated temperature-pressure profile along with altitude, and metallicity.
    \item We use the resultant chemical abundances from the equilibrium chemistry run as initial conditions and run the disequilibrium chemistry model until it reaches a steady-state.
    \item We use the respective chemical abundances outputs from the equilibrium and disequilibrium chemistry runs as inputs into the radiative transfer model, \texttt{Pyrat Bay}, and produce high-resolution emission spectra.
    \item We use the high-resolution spectra as inputs into the JWST simulator, \texttt{PandExo} to produce simulated JWST spectra.
\end{enumerate}
The specifics of each of these steps is explained in the following sections. 

\subsection{Temperature-Pressure Profiles}
\label{sec:tp}
For our study, we use 1D globally averaged temperature-pressure (TP) profiles for the sake of inter-comparison with the profile from the well-studied planet of HD 189733b. The TP profile for HD 189733b is taken from \citet{Moses2011} (also seen in \citealt{Venot2012}). This profile is derived from both a 1D globally-averaged profile from \citet{Fortney2006, Fortney2010} and 3D global circulation models (GCM) from \citet{Showman2009}. (See \citealt{Moses2011} for details.) Additionally, \citet{Moses2011} heats the upper atmosphere of this TP profile, describing the addition of this heating as an  ``\textit{ad hoc}'' procedure, adapted from the results of \citet{Yelle2004, GarciaMunoz2007, Koskinen2010}. \citet{Koskinen2010} models of the upper atmosphere for HD 209458b, and calculates the lower boundary of its thermosphere as $\sim$0.1$\mu$bar. \citet{Moses2011} employs this same pressure boundary for the TP profile of HD 189733b despite the differences in gravity (9.4 m s$^{-2}$ for HD 209458b, 22.8 m s$^{-2}$ for HD 189733b), as the gravity affects the scale height and ultimately the location of this boundary. It is important to note that the foundational work on aeronomy in giant gas exoplanets was conducted in \citet{Yelle2004} on globally-averaged TP profiles, citing the use of 1D in lieu of 3D models for the modelling of Titan's atmosphere as a valid approach \citep{MullerWodarg2000}. For the sake of consistency, we apply the same heating procedure as \citet{Moses2011} to the profiles of  WASP-80b and GJ 436b. The TP profiles for WASP-80b and GJ 436b are generated by the non-grey analytical model of \citet{Parmentierderivation} assuming global redistribution ($\alpha=0.25$) as 3D GCM results for these planets are unavailable. These temperature-pressure profiles do not include any upper atmosphere heating so we apply the same heating procedure as \citet{Moses2011} to the profiles of WASP-80b and GJ436b again ignoring the differences in gravity (15.8 m s$^{-2}$ for WASP-80b, 12.6 m s$^{-2}$ for GJ436b).  The addition of this heating is achieved by cross-correlation to the TP profile of HD 189733b from \citet{Moses2011} for pressures less than 0.1 $\mu$bar. The TP profiles used in our study are not calculated self-consistently. 

For the calculation of scale height, we assume hydrostatic equilibrium, using a mean molecular weight of 2.3 for HD 189733b and WASP-80b, and 4.6 for GJ 436b \citep{Line2016}.

We use the same eddy diffusion profile originally calculated in \citet{Moses2011} for all three planets. \citet{Moses2011} calculates a globally-averaged eddy diffusion profile from the 3D results of \citet{Showman2009}. Little is known about eddy diffusion (as mentioned in \citealt{Moses2011})  thus, we choose to employ this calculated profile rather than a constant in an attempt to employ the state-of-art in eddy diffusion. Although, the use of different eddy diffusion profiles for different planets could change our results, we instead aim to study the impacts of radius, temperature, and metallicity on the observability of disequilibrium with JWST. Thus, to study these parameters, we employ the same eddy diffusion profile \citep{Moses2011} throughout our study to isolate these parameters. The sensitivity of eddy diffusion on calculated chemical composition is discussed in \citet{Moses2011}.

\subsection{Chemistry models and Bulk Composition}
\label{sec:chemmod}
\begin{figure}[tb]
\centering
\includegraphics[width=\linewidth, clip]{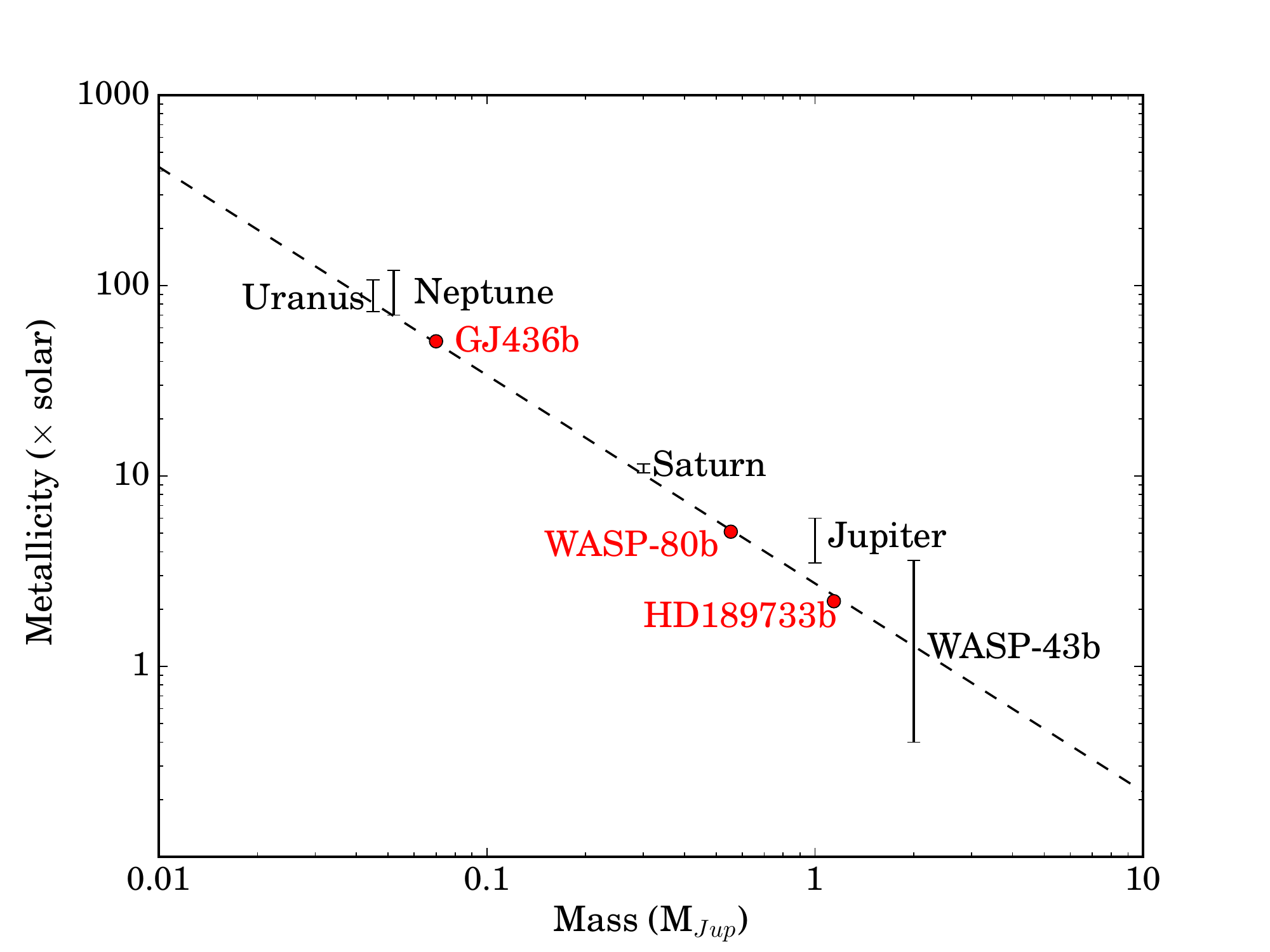}
\caption{The planets analyzed in this paper used metallicites calculated from the solar system derived metallicity assumption seen in \citet{Kreidberg2014} (dashed line); Figure adapted directly from \citet{Kreidberg2014} with planets in this study in red.} 
\label{fig:metallicity}
\end{figure}

For bulk composition,  we investigate the effect of metallicity by comparing two different metallicities. We compare solar metallicity and the metallicity derived from the mass-metallicity relationship described in \citet{Kreidberg2014} (now to be referred to as Kr14), where we scale from the solar abundance from \citet{Asplund}; see Figure \ref{fig:metallicity} and Table \ref{table:planetsumm}. We adopt the Kr14 metallicity for the `enriched' case as a unifying convention as the metallicity has been explored for HD 189733b and GJ 436b, but not for WASP-80b. We do not take into account the uncertainty in the Kreidberg-derived metallicty, but simply use metallicites that fall directly on the Kreidberg-fitted line. We input both solar and Kr14 metallicities into the equilibrium chemistry model. This elemental abundance is then propagated through to the disequilibrium chemistry model.

We use two chemistry models--- one that models thermochemical equilibrium which we call equilibrium for short, and the other which models chemical kinetics with photochemistry which we call disequilibrium. Our equilibrium chemistry model minimizes the Gibbs energy following the algorithm of \citet{GordonMcBride1994}, adopting the same thermochemical data as \citet{Venot2012} in the form of NASA polynomial coefficients \citep{McBrideGordonReno:1993a}. Our disequilibrium chemistry model is a classical one-dimensional model of planetary atmospheres in which thermochemical kinetics, vertical transport, and photochemistry are taken into account. A nonlinear system of first-order ordinary differential equations (ODE) discretized along a vertical column of atmosphere is integrated as a function of time. It starts from an initial composition, using a backward differentiation formula implicit method for stiff problems implemented in the Fortran solver DLSODES within the ODEPACK package \citep{Hindemarsh1983, RadhaandHindemarsh1993}. Thermochemical kinetics rely on a chemical network consisting of 102 neutral species composed of C, H, N, and O linked by 1918 chemical reactions, that has been validated in the area of combustion chemistry by numerous experiments over the 300-2500 K temperature range and the 0.01-100 bar pressure regime. This has been found to be suitable to model the atmospheres of hot Jupiters  \citep[see][]{Venot2012}. Most reactions are reversed with their rate constants fulfilling detailed balance to ensure that thermochemical equilibrium is achieved at sufficiently long times in the absence of any disequilibrium processes (e.g. photochemistry and vertical transport). 

Absorption and photodissociation cross sections are also taken from \citet{Venot2012}. The different incident UV fluxes adopted follow the stellar models from \citet{Rugheimer2013} and \citet{Rugheimer2015}, detailed in Table \ref{table:planetsumm}. The resulting UV fields in the atmospheres are calculated in spherical geometry for a 45$^{\circ}$ zenith angle after attenuation by both absorption and Rayleigh scattering (through a two-ray iterative algorithm, \citep{Isaksen1977}. The ODE system is integrated from a chemical composition calculated at thermochemical equilibrium, with different elemental abundances \citep{Asplund, Kreidberg2014} and the temperature-pressure profile, discussed in Section \S\ref{sec:tp}.

\begin{table*}[ht]
\centering
\caption{\label{table:planetsumm} Summary of Planets}
\scalebox{0.9}{
\begin{tabular}{lccccccccc}
\hline
\hline
Planet      &  $R_\mathrm{p}$ ($R_\mathrm{Jupiter}$) & $M_\mathrm{p}$ ($M_\mathrm{Jupiter}$) & Kreidberg Metallicity & $T_\mathrm{eff}$ (K) & $T_\mathrm{star}$ (K) & Stellar Type &  Stellar Age         & J magnitude & Stellar Model \\
\hline
HD 189733b  &  1.138 & 1.142 &  2 & 1191         & 4875           & K1-K2        & $>$6 x 10$^{8}$ years  & 6.07 & Eps Eridani\footnote{from \citet{Rugheimer2013} \label{ref:Rug13} } \\
WASP-80b    &  0.952 & 0.554 &  5 & 814        & 4150           & K7V          &  unknown             & 9.22 & BY Dra\footref{ref:Rug13} \\
GJ436b      & 0.38  & 0.007 & 50 &712        & 3684           & M2.5         & 6.0 ($\pm$ 5.0) Gyr  & 6.99 & GJ 436\footnote{from \citet{Rugheimer2015} } \\
\hline
\end{tabular}
}
\end{table*}

\subsection{Spectra Generator}
\label{subsec:transit}

To model the planetary emission spectra, we use the Python Radiative-transfer in a Bayesian framework package, \texttt{Pyrat Bay}, \citep{CubillosEtal2017apjPyratBay}.  \texttt{Pyrat Bay} is an open-source, reproducible package\footnote{\href{https://github.com/pcubillos/pyratbay} {https://github.com/pcubillos/pyratbay}.}, which is updated from the Bayesian Atmospheric Radiative Transfer package \citep{Cubillos2016phdThesis, Blecic2016phdThesis}.

\texttt{Pyrat Bay} solves the one-dimensional radiative-transfer equation, using the input atmospheric models (pressure, temperature, altitude, and abundances profiles) assuming a plane-parallel geometry.  The code computes the emergent intensity spectra for a range of angles with respect to the normal.  It then integrates the contribution from the different angles to produce the day-side flux spectrum.  The radiative-transfer equation considers both absorption and emission from the atmosphere, adopting local-thermodynamic equilibrium to approximate the source function as
the Planck function.

We consider the opacities from the expected major species in gas-giant planets.  We incorporate molecular opacities from the `line-by-line' HITRAN and HITEMP databases \citep{Rothman2013JqsrtHITRAN, RothmanEtal2010jqsrtHITEMP}, see Table \ref{table:opacitysources}; the collision-induced absorption opacities for H$_{2}$--H$_{2}$ \citep{BorysowEtal2001jqsrtH2H2highT, Borysow2002jqsrtH2H2lowT} and H$_{2}$--He \citep{BorysowEtal1988apjH2HeRT,BorysowEtal1989apjH2HeRVRT, BorysowFrommhold1989apjH2HeOvertones};
and Rayleigh scattering opacity \citep{LecavelierDesEtangsEtal2008aaRayleighHD189}. Our calculation is done at a resolution of 1 cm$^{-1}$.

All inputs and outputs for our \texttt{Pyrat Bay} runs can be found at \href{https://github.com/sdb05c/FiLeS\_EQNEQ} {https://github.com/sdb05c/FiLeS\_EQNEQ}.

\begin{table}[ht]
\centering
\caption{\label{table:opacitysources} Molecular Opacity Sources}
\begin{tabular}{llll}
\hline
\hline
Molecule     & Database & Spectral Coverage (cm${-1}$ & Number of Lines \\
\hline
H$_{2}$O      & HITEMP   & 0-30,000                    & 114,241,164     \\
CO$_{2}$      & HITEMP   & 258-9,648                   & 11,193,608      \\
CO           & HITEMP   & 3-8,465                     & 113,631         \\
NO           & HITEMP   & 0-9,274                     & 115,610         \\
N$_{2}$O      & HITRAN   & 0-7,797                     & 47,843          \\
CH$_{4}$      & HITRAN   & 0-9,200                     & 290,091         \\
O$_{2}$       & HITRAN   & 0-15,928                    & 6,428           \\
NO           & HITRAN   & 0-9,274                     & 105,079          \\ 
NO$_{2}$      & HITRAN   & 0-3,075                     & 104,223          \\
NH$_{3}$      & HITRAN   & 0-5295                      & 29,084           \\
H$_{2}$CO     & HITRAN   & 0-3,100                     & 37,050           \\
HCN          & HITRAN   & 0-3,424                     & 4,253            \\
H$_{2}$O$_{2}$& HITRAN   & 0-1,731                     & 126,983          \\
C$_{2}$H$_{2}$& HITRAN   & 604-9,890                   & 11,340           \\
C$_{2}$H$_{6}$& HITRAN   & 706-3,001                   & 22,402           \\
C$_{2}$H$_{4}$& HITRAN   & 701-3,243                   & 18,378           \\
\hline

\hline
\end{tabular}
\end{table}

\subsection{JWST Simulator}
We use \texttt{PandExo} \citep{Batalha2017PandExo} \footnote{\href{https://github.com/natashabatalha/PandExo} {https://github.com/natashabatalha/PandExo}} to simulate secondary eclipse observations of planets akin to HD 189733b, WASP-80b, and GJ 436b. We chose to simulate spectra using the Near-InfraRed Imager and Slitless Spectrograph's Single Object Slitless Spectroscopy mode (NIRISS SOSS) from 0.6 to 2.8 $\mu$m (R=700), Near InfraRed Spectrograph's Grism 395 Medium Resolution mode (NIRSpec G395M) from 2.9 to 5 $\mu$m (R=1000), and the Mid-InfraRed Instrument's Low Resolution Spectrometer (MIRI LRS) from 5 to 14 $\mu$m (R=100).

These observing modes were chosen because they offer nearly full wavelength coverage from 1-12 $\micron$. NIRCam Grism with F322W2 and F444W2 could theoretically be used instead of NIRSpec G395M. Using these two NIRCam modes would decrease the noise in the 4-5 $\micron$ range but at the cost of 2$\times$ the observing time. The NIRSpec Prism, which offers wavelength coverage from 1-5 $\micron$ cannot be used for HD 189733,  WASP-80, or GJ 436 because all three planet systems are past the Prism's saturation limit. 

In the calculation of noise, \texttt{PandExo} does not assume a systematic noise floor. \citet{Greene2016} have suggested, based on previous observations with HST and Spitzer, that the noise floor for JWST might be 20-50 ppm. This will not be truly known until well after commissioning of the telescope. The assumption to not include a noise floor does not change the conclusions presented here.

\texttt{PandExo} uses the Phoenix Stellar Atlas models \citep{Husser2013}. Each transit observation consists of equal time spent in and out of transit (total time = 2$\times$ transit duration). HD 189733 and GJ 436 are near the saturation limits of the instruments and their observations contain 2 groups/integration, which corresponds to a duty cycle of 0.33. WASP-80, with a J=9.22, is well away from the saturation limit of NIRISS, NIRSpec G395M and MIRI and therefore can be observed with a much higher efficiency (14 groups/integration and duty cycle of 0.86).

\section{Results \& Discussion} 
\label{sec:results}
\begin{figure*}[htb]
\centering
\includegraphics[width=0.3\linewidth, clip]{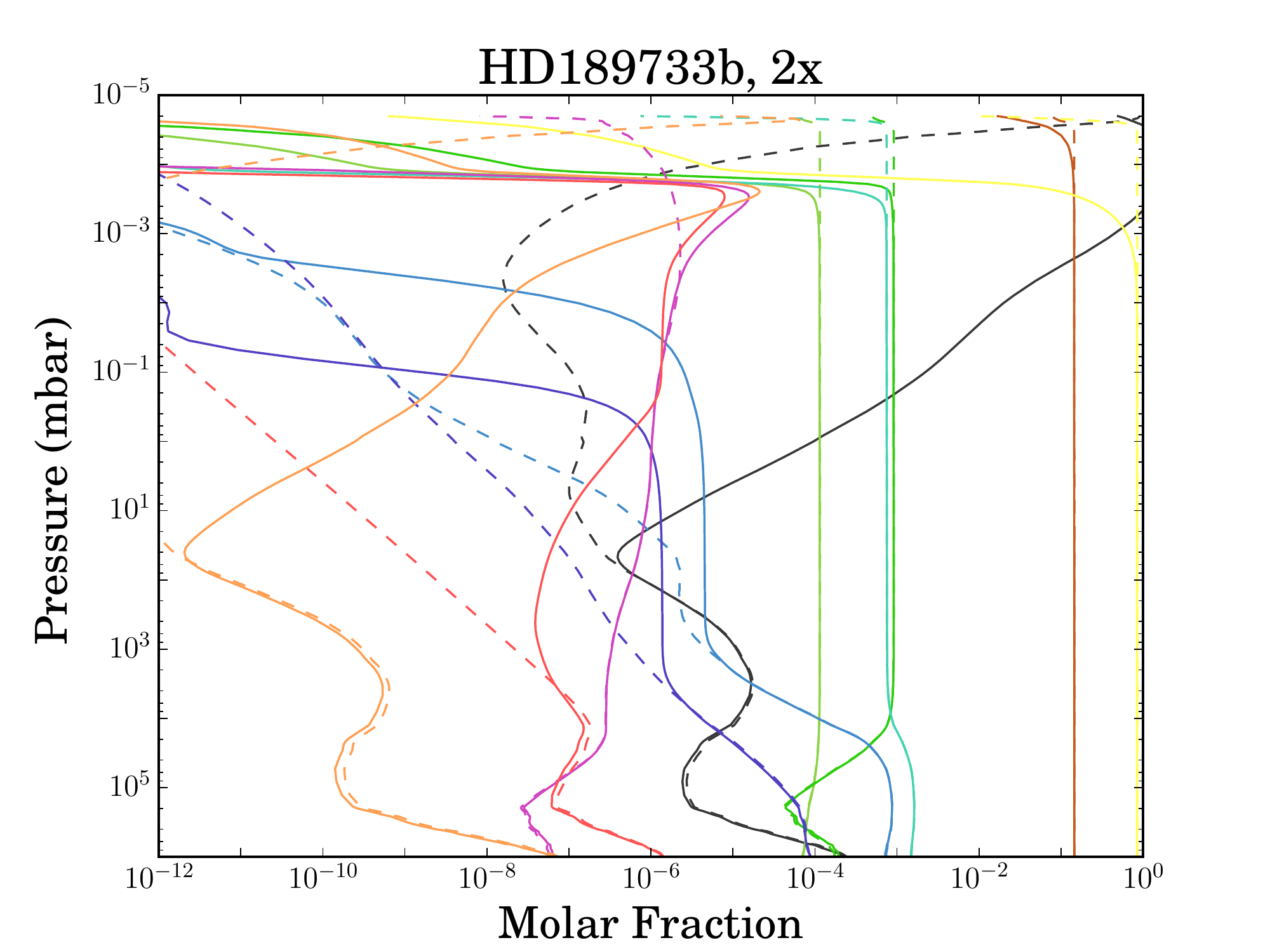}
\includegraphics[width=0.3\linewidth, clip]{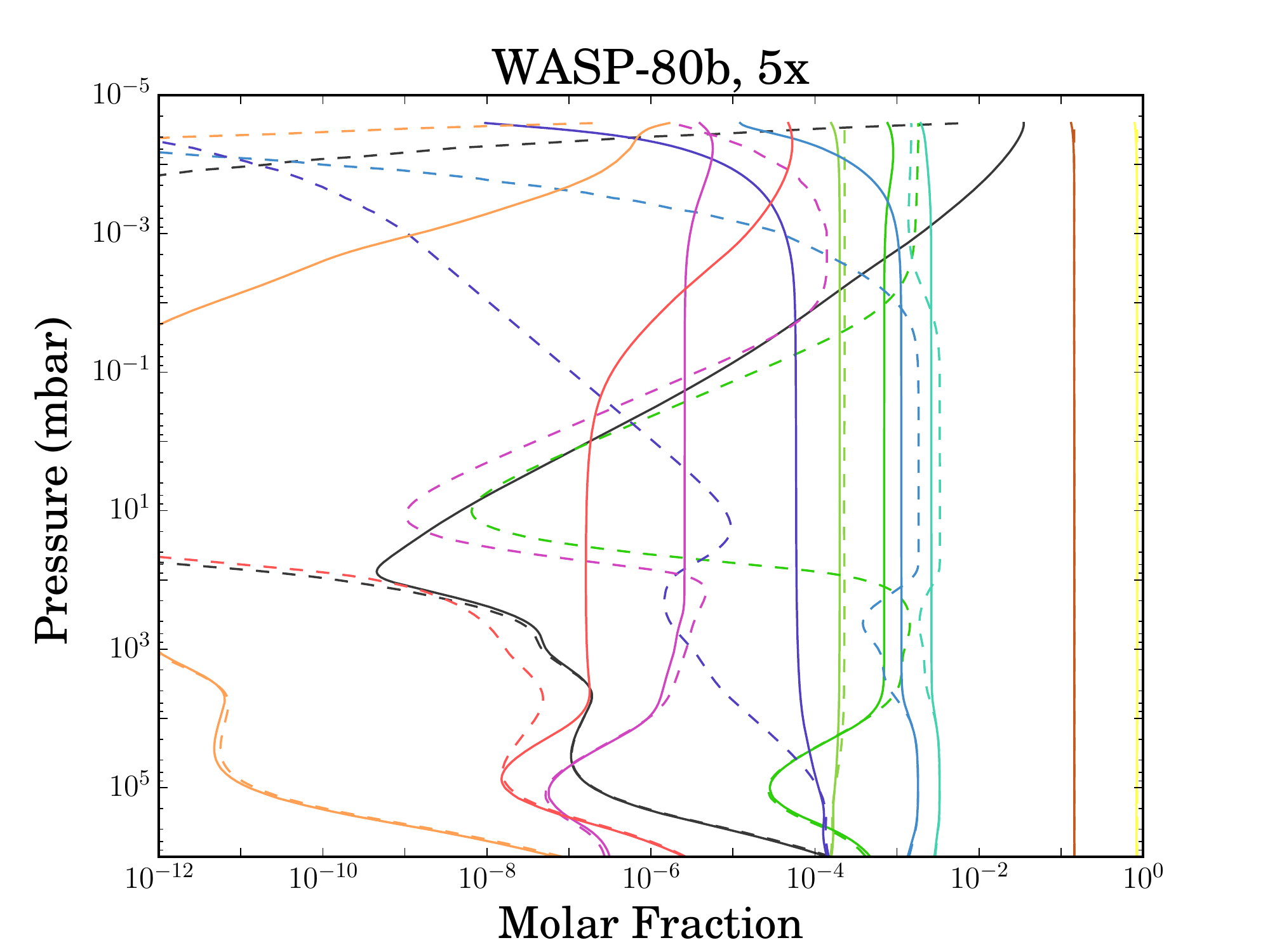}
\includegraphics[width=0.3\linewidth, clip]{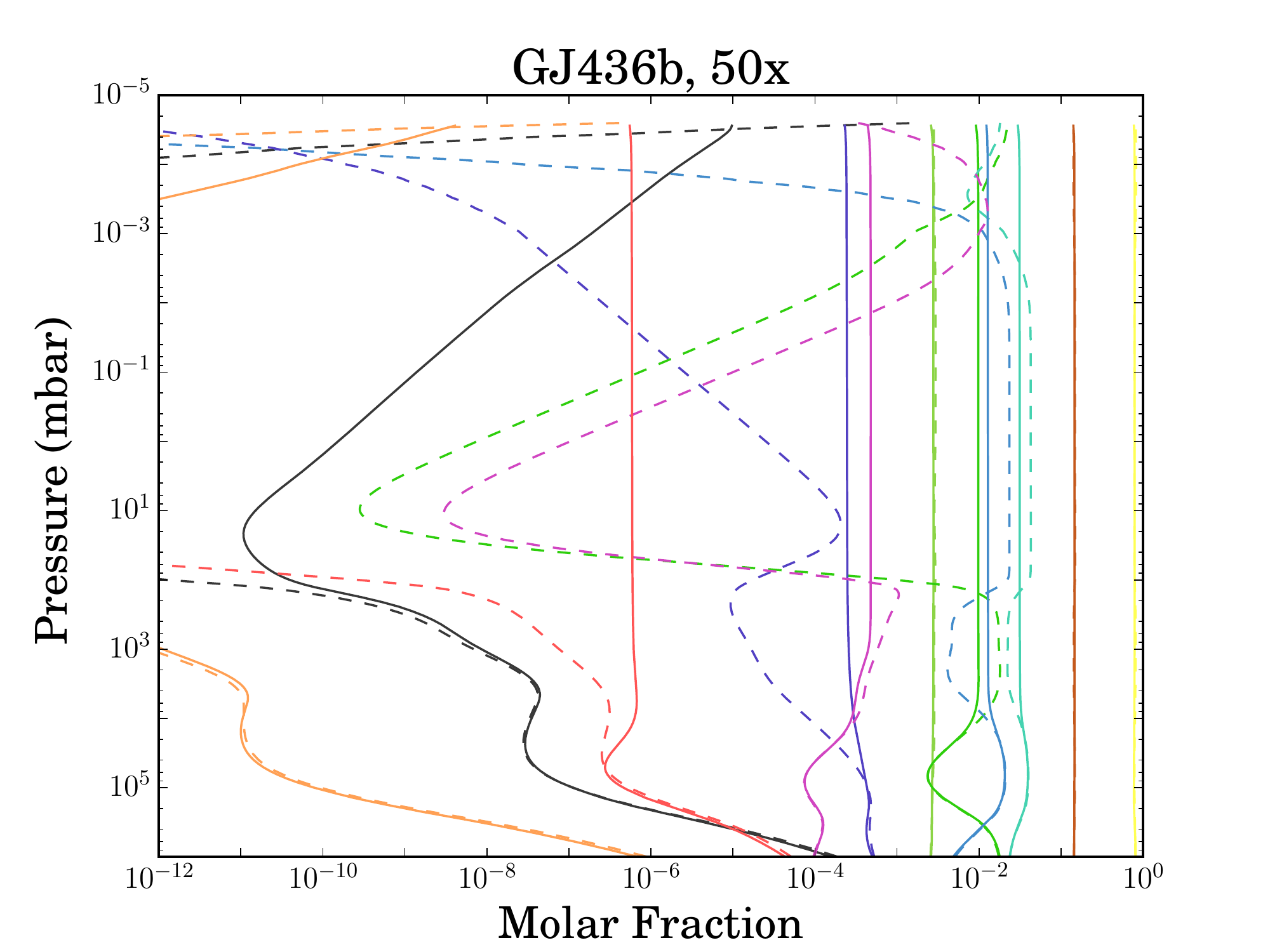}
\includegraphics[width=0.08\linewidth, clip]{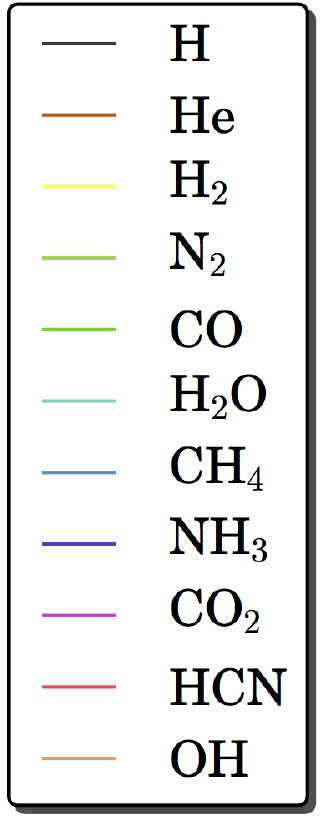}

\includegraphics[width=0.3\linewidth, clip]{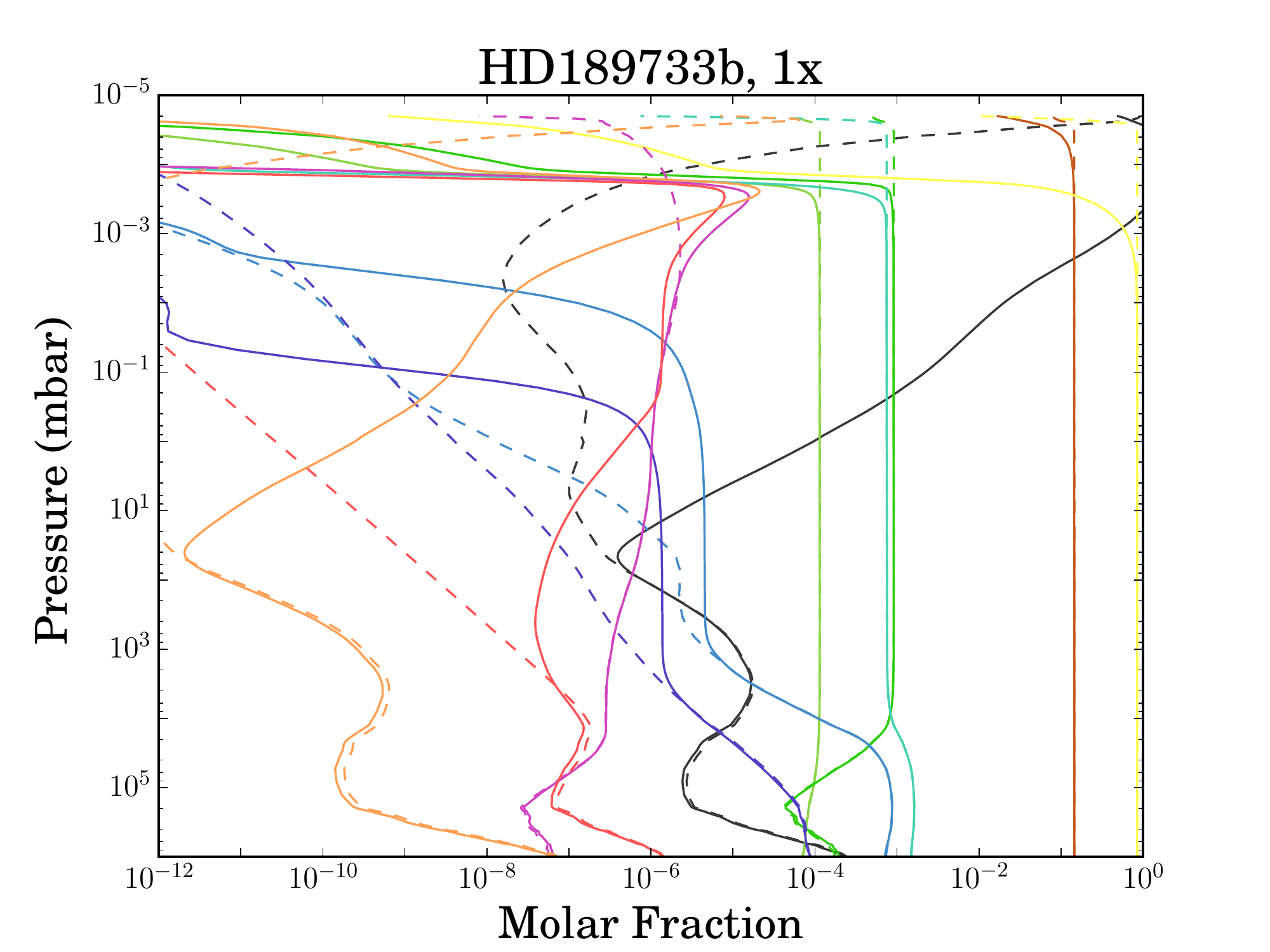}
\includegraphics[width=0.3\linewidth, clip]{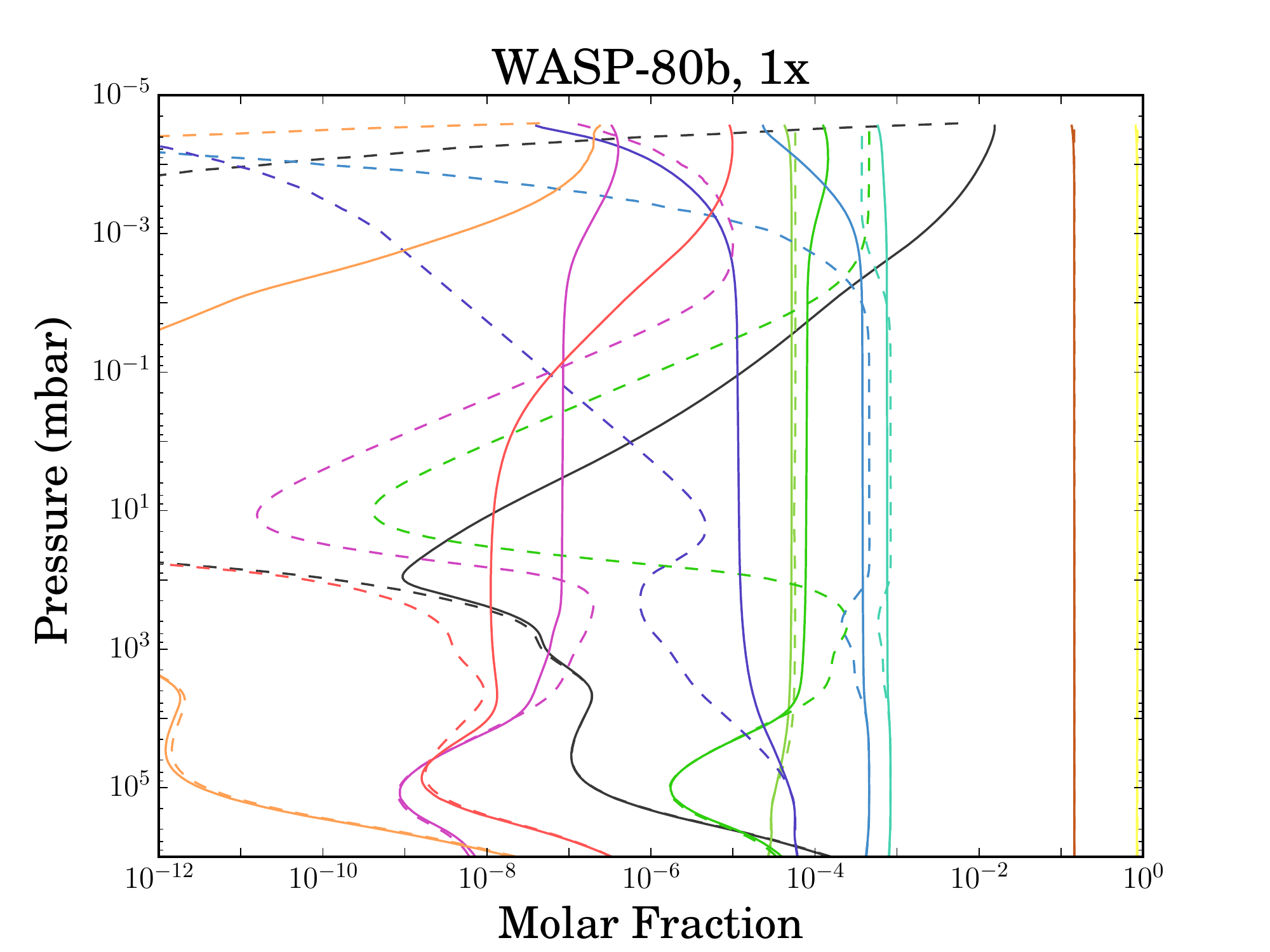}
\includegraphics[width=0.3\linewidth, clip]{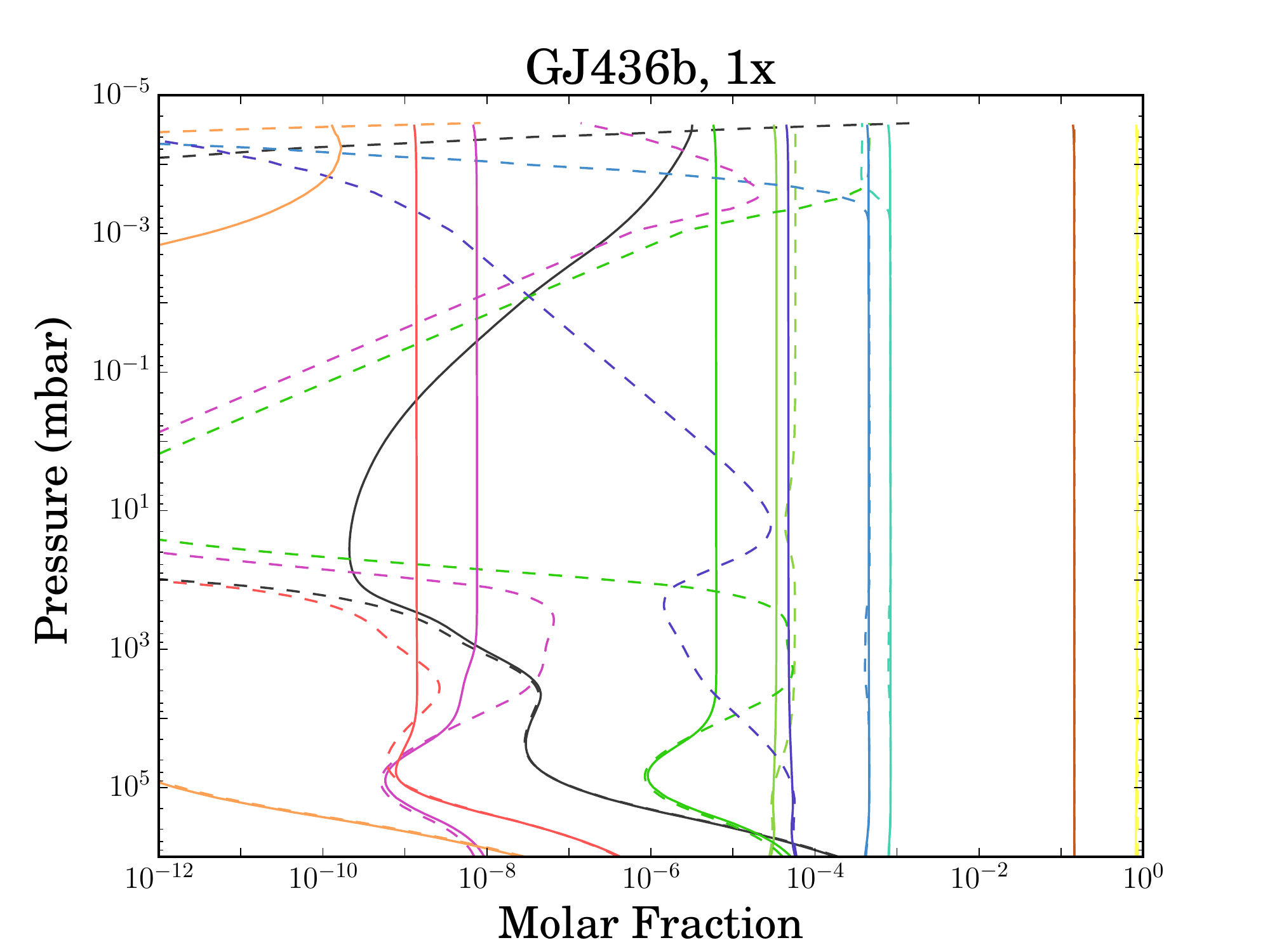}
\includegraphics[width=0.08\linewidth, clip]{legend.pdf}

\caption{Equilibrium (\textit{dashed lines}) versus disequilibrium (\textit{solid lines}) chemical abundances for solar metallicity (\textit{top row}) and Kr14 metallicity (\textit{bottom row}) for planets like HD 189733b, WASP-80b, and GJ 436b}
\label{fig:chem_majspec}
\end{figure*}

We examine the impacts of metallicity, temperature and radius on the observability of disequilibrium chemistry in the atmospheres of our model planets with JWST.  First, we present results showing the divergences from chemical equilibrium assuming either solar metallicity (1$\times$), or a Kr14 metallicity. We examine the resultant chemical abundances seen in Figure \ref{fig:chem_majspec}. As previously shown in \citet{Moses2011} and \citet{Venot2012} for a solar metallicity, a lower equilibrium temperature leads to a greater divergence from chemical equilibrium. Our results confirm this as well. Overall, we see more divergence from equilibrium in our GJ 436b-like planet than in our HD 189733b-like planet. This is because the lower the ambient temperature (and pressure), the slower the rate at which reactions will take place thus causing the atmosphere to be farther away from equilibrium. When comparing close metallicities for the HD 189733b-like planet (1$\times$ and 2$\times$), the computed abundances are nearly identical. See \citet{Venot2012} for a complete discussion on the chemical divergences from equilibrium for each species. Our WASP-80b-like planet, which has a equilibrium temperature 400K colder than HD 189733b, shows larger divergences from chemical equilibrium. For both 1$\times$ and 5$\times$ solar metallicity, we see similar functions of abundance versus pressure, however for the 5$\times$ solar metallicity case, as there are more `metals' available, so overall, molecules are in higher fraction. CO, CO$_{2}$, and NH$_{3}$ show the most divergence, and CH$_{4}$ has large divergence at low pressure. For the GJ 436b-like planet, it is similar to the WASP-80b-like planet in that we see again the most significant divergences in CO, CO$_{2}$, and NH$_{3}$, and large divergence for CH$_{4}$ at low pressures. For the 50$\times$ solar case, molecules are again in higher fraction as the availability of `metals' is higher than the solar metallicity case. We see a larger magnitude of divergence for the solar metallicity case than 50$\times$ solar metallicity case. 

It is also important to mention the quenching in each of these cases. We can see more species being clearly quenched  in the WASP-80b-like and GJ 436b-like planet cases--- CO, H$_{2}$O, CH$_{4}$, NH$_{3}$, CO$_{2}$, HCN---  than in the HD 189733b-like planet cases. In the HD 189733b-like planet case, we only see CO and H$_{2}$O being clearly quenched. However the quenching behavior of HCN in the HD 189733b-like cases is similar to the HCN quenching behavior in the WASP-80b-like cases. Additionally, it is important to note the very slight divergence from equilibrium water undergoes in the Kr14 metallicity cases. As the metallicity increases and the temperature decreases, we see more divergence. 

\begin{figure*}
\centering
\includegraphics[width=\linewidth]{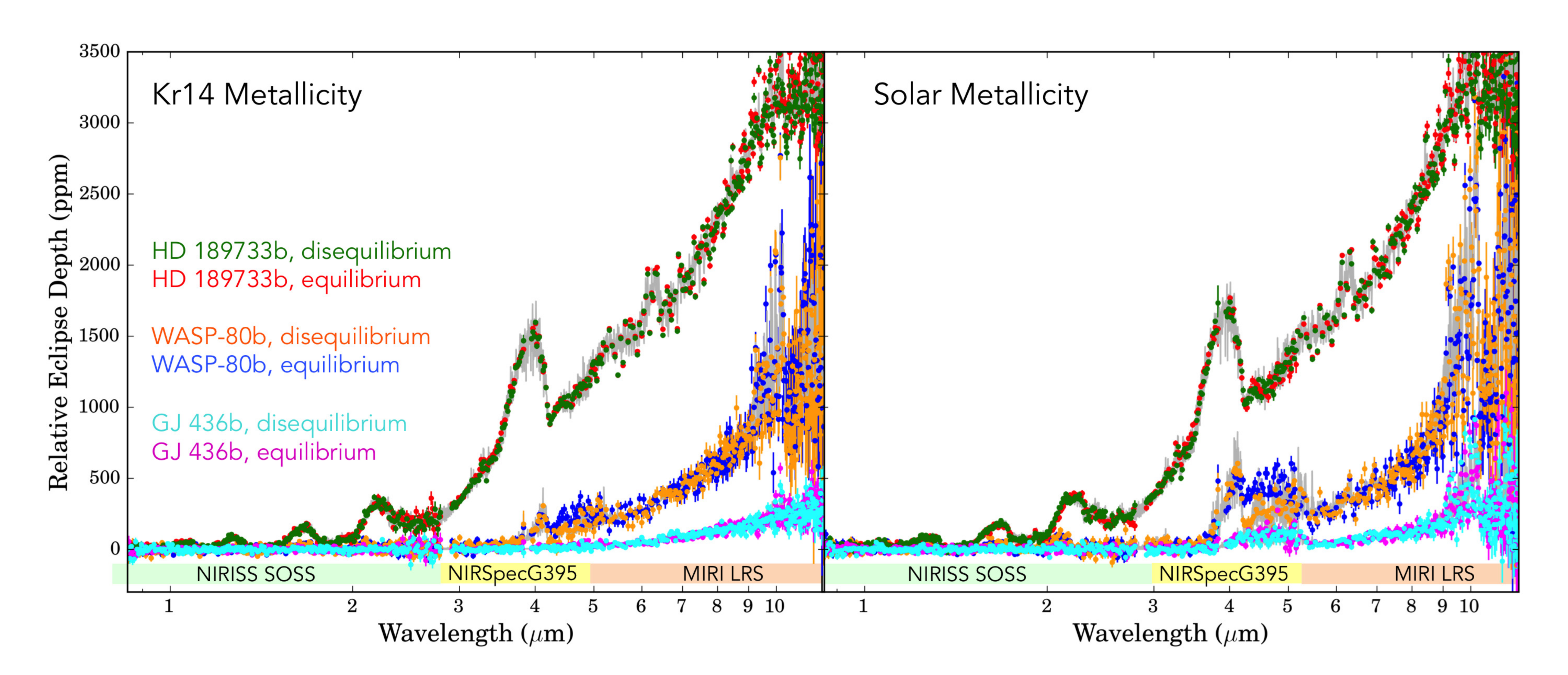}
\caption{JWST simulated spectra for the three planets for 5 eclipse events at an R=100. HD 189733b, equilibrium and disequilibrium, respectively in red and green, WASP-80b, in blue and orange, GJ 436b, in magenta and cyan; \textit{left} Kreidberg-derived (Kr14) metallicity, and \textit{right} solar metallicity for the three JWST instruments. The disequilibrium model spectra is under-plotted in light grey.}
\label{fig:JWST_spec}
\end{figure*}

Using the equillibrium and non-equillibrium abundances shown in Figure \ref{fig:chem_majspec} as inputs, we examine the impacts of metallicity, radius, and temperature on the observability of non-equilibrium chemistry processes in the atmospheres of our planets with JWST's NIRISS SOSS, NIRSpec G395M, and MIRI LRS (Figure \ref{fig:JWST_spec}). Our criteria for observability is defined as a difference in single-bin eclipse depth of greater than 60 ppm in the NIR and 150 ppm in the MIR; this would be a 3-sigma result based on expected noise floors for the JWST instruments \citep{Greene2016}. In order to distinguish between model spectra at a level above the noise contribution for each bin, we model the spectra seen in Figure \ref{fig:JWST_spec} assuming a total of five eclipse events. Divergence between equilibrium and disequilibrium chemistry becomes observable from 4 to 5 $\mu$m using the NIRSpec G395M. We can not disentangle equilibrium from disequilibrium chemistry using NIRISS SOSS or MIRI LRS for any of our selected cases.

 As seen in Figure \ref{fig:chem_majspec}, the largest chemical differences between equilibrium and disequilibrium compositions are in the GJ 436b-like planet, the coolest in of our study. We expected these differences to be the most noticeable by JWST but due to the radius of this planet, these differences are not as observable as those for the larger and slightly warmer WASP-80b-like planet. The magnitude of the observed flux and the magnitude of the difference between spectra from equilibrium and disequilibrium models is greater for solar metallicity planets than enriched planets.  The difference between equilibrium and disequilibrium chemistry becomes observable from 4 to 5 $\mu$m for both metallicities for the WASP-80b-like planet, and barely observable for the GJ 436b-like planet at solar metallicity, but not observable at 50$\times$ solar metallicity. The average difference from 4 to 5 $\mu$m for a GJ 436b-like planet is 30 ppm for solar metallicity, which is only 1.5$\sigma$ result according to noise predictions made in \citet{Greene2016}. The average difference from 4 to 5 $\mu$m for a WASP-80b-like planet is 60 ppm for 5$\times$ solar metallicity, and 130 ppm for solar metallicity. As metallicity increases, the intensity of spectral features decreases. Thus the sweet spot from 4 to 5 $\mu$m depends also on metallicity. The difference in radii between WASP-80b and GJ 436b is $\sim$86\% (calculated from Table \ref{table:planetsumm}). We see no discernible differences for our HD 189733b-like planet at both solar and 2$\times$ solar metallicities. The difference in radius between HD 189733b and WASP-80b is $\sim$18\% but the difference in temperature is $\sim$40\% (calculated from Table \ref{table:planetsumm}). Thus, this sweet spot is made up of the intersection of temperature, radius, and metallicity. It is important to note that these results are highly dependent on the fixed inputs presented and could or could not be discernible under different conditions.

\begin{figure*}
\centering
\includegraphics[width=0.45\linewidth]{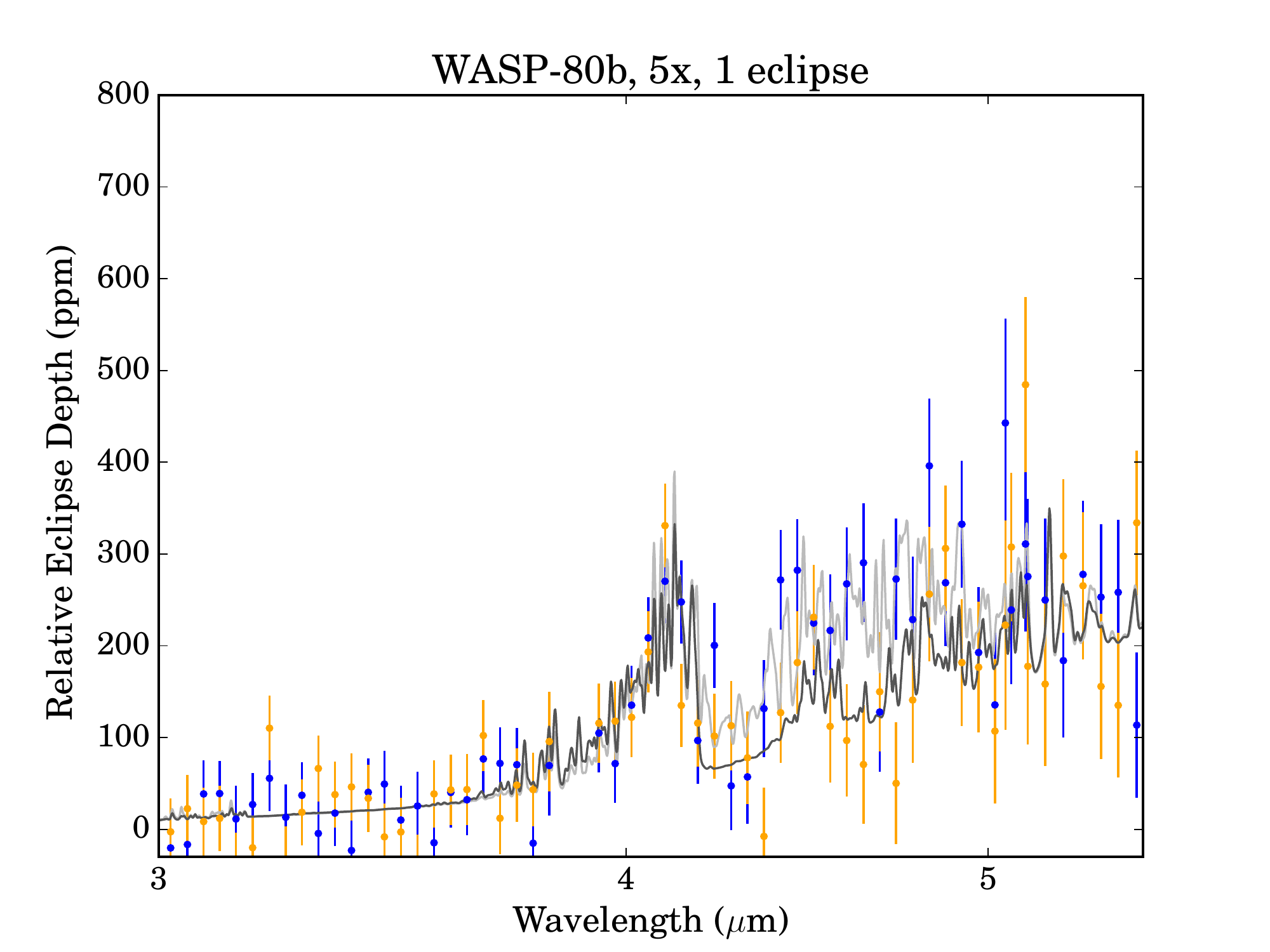}
\includegraphics[width=0.45\linewidth]{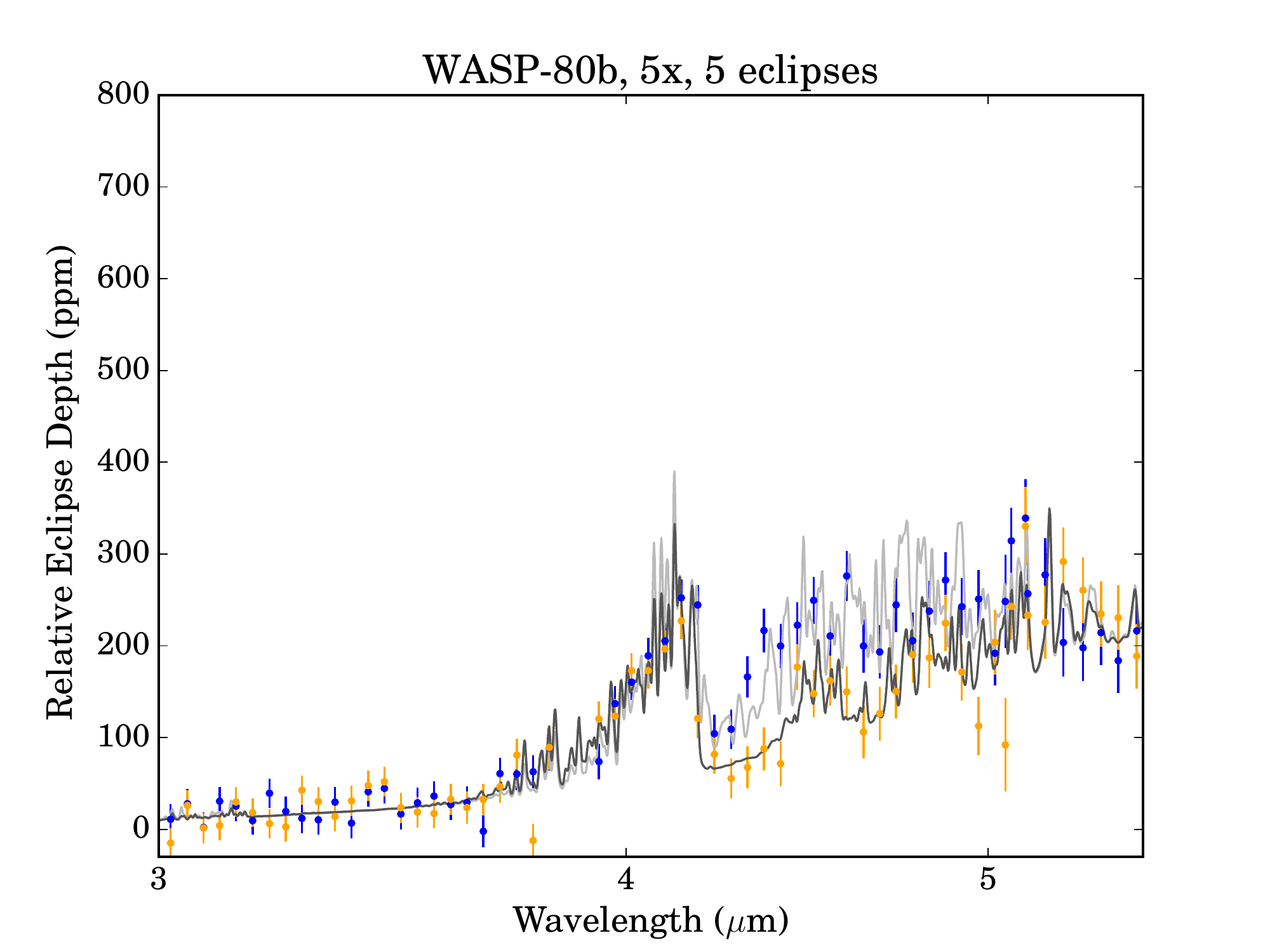}
\includegraphics[width=0.45\linewidth]{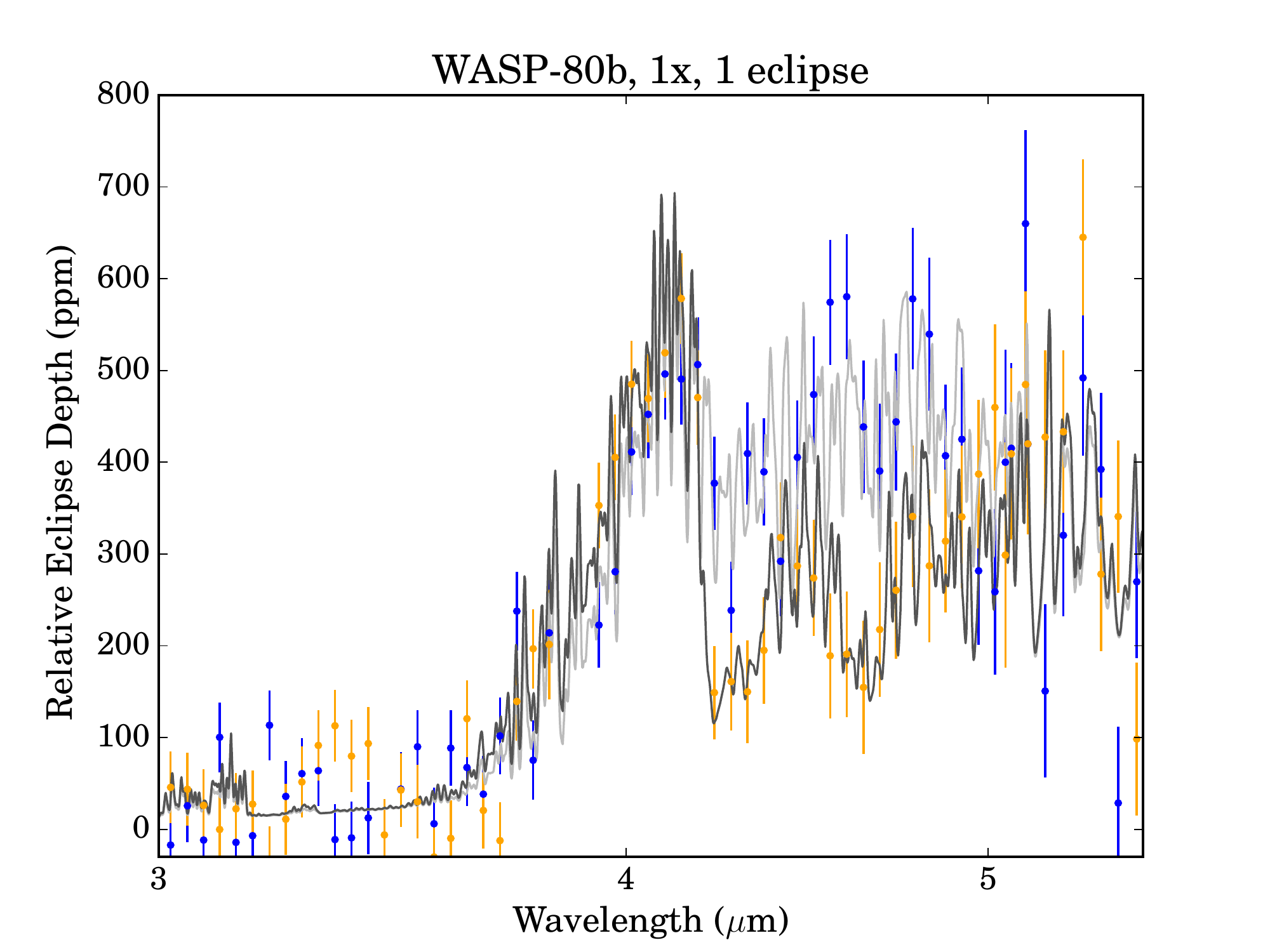}
\includegraphics[width=0.45\linewidth]{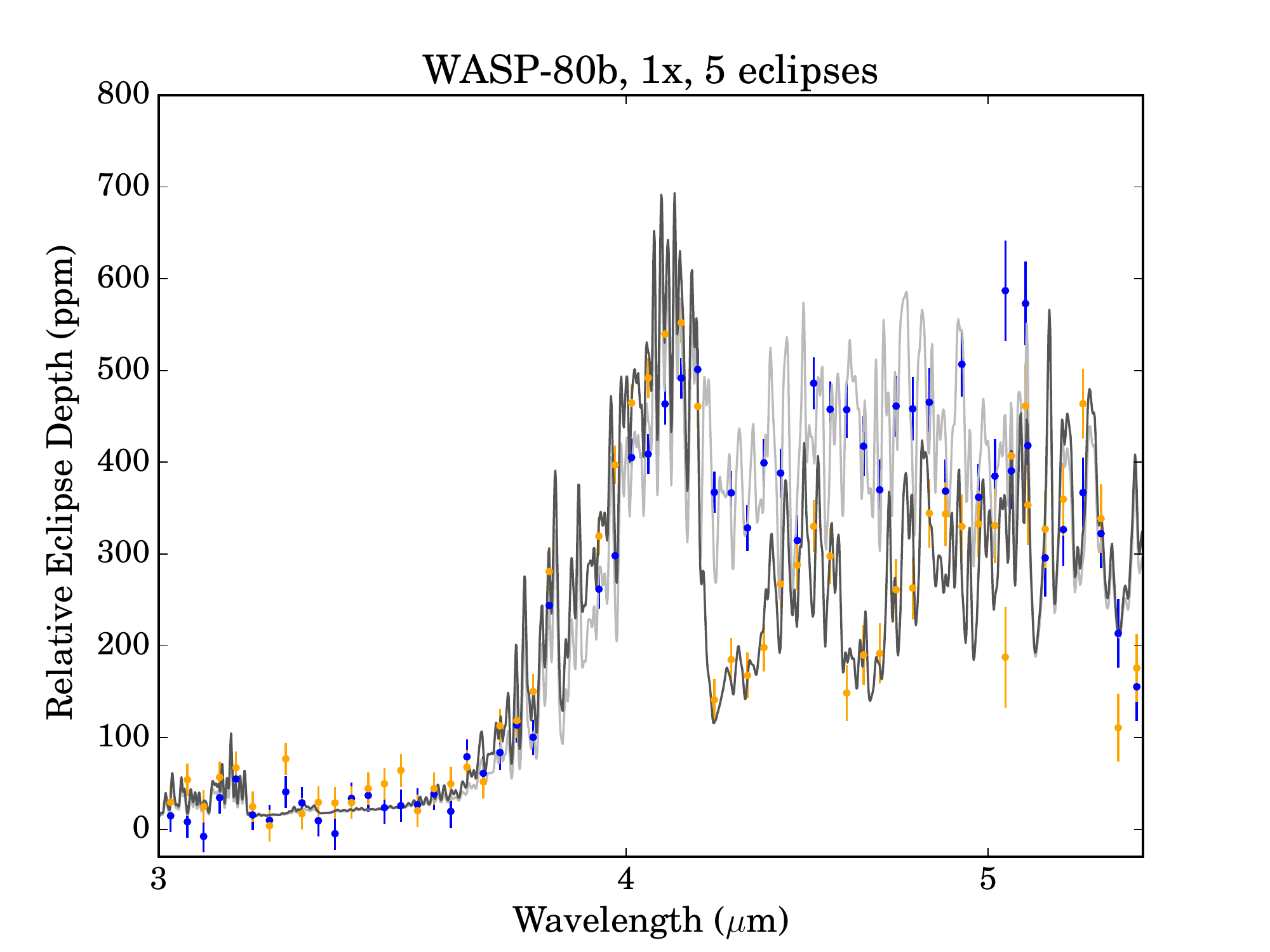}
\caption{Zoomed region of JWST simulated spectra for the WASP-80b-like planet to highlight the observable differences seen from 4 to 5 $\mu$m at R=50. Equilibrium chemistry points in blue, high resolution model in light grey, and disequilibrium points in yellow, high resolution spectra in darker grey. Modelled with the \textit{row 1}: Kr14 metallicity (5$\times$), and \textit{row 2}: solar metallicity, for \textit{left column}: 1 eclipse event  and \textit{right column}: 5 eclipse events . }
\label{fig:JWST_speczoom_WASP80b}
\end{figure*}

\begin{figure*}
\centering
\includegraphics[width=0.45\linewidth]{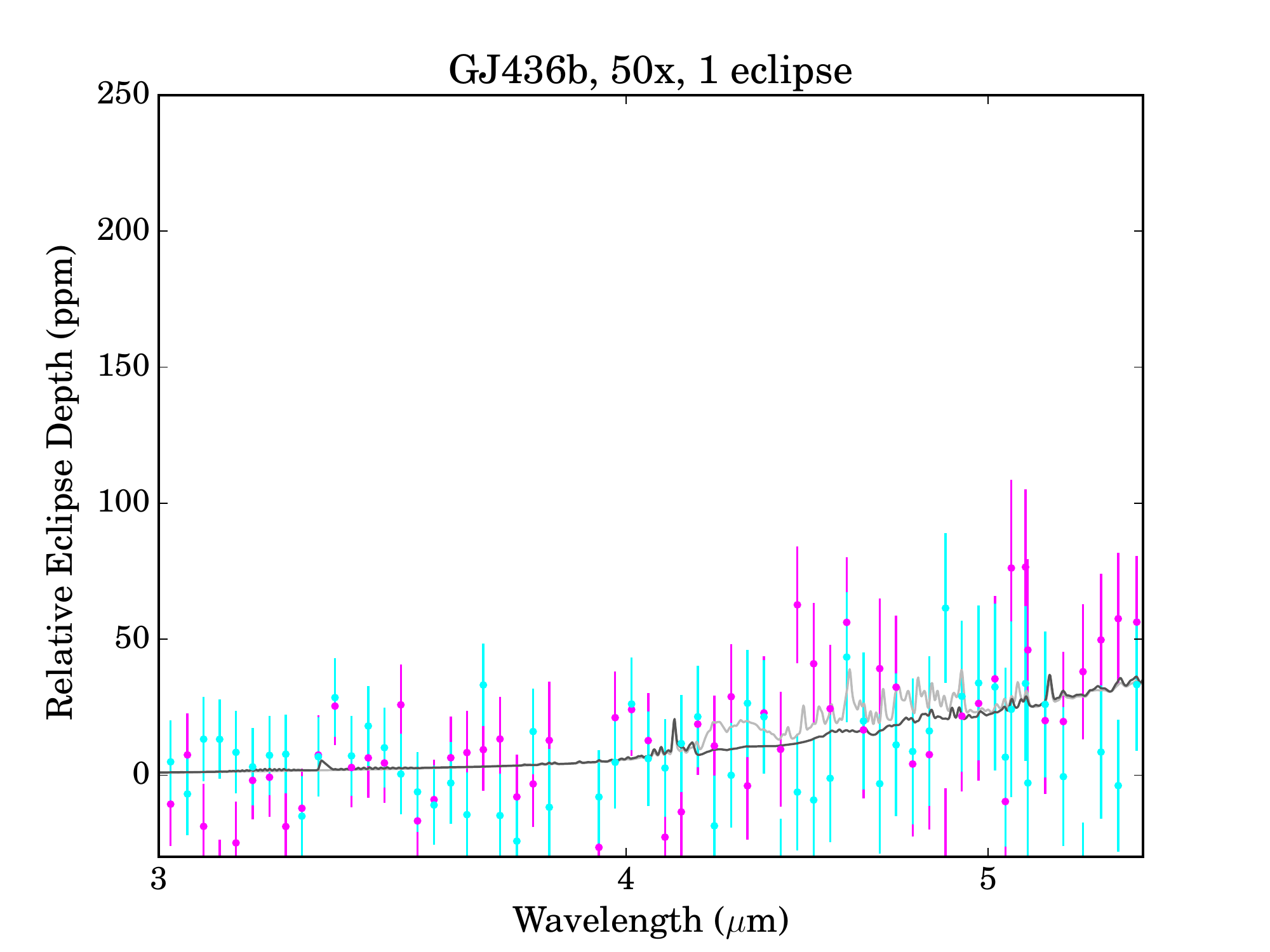}
\includegraphics[width=0.45\linewidth]{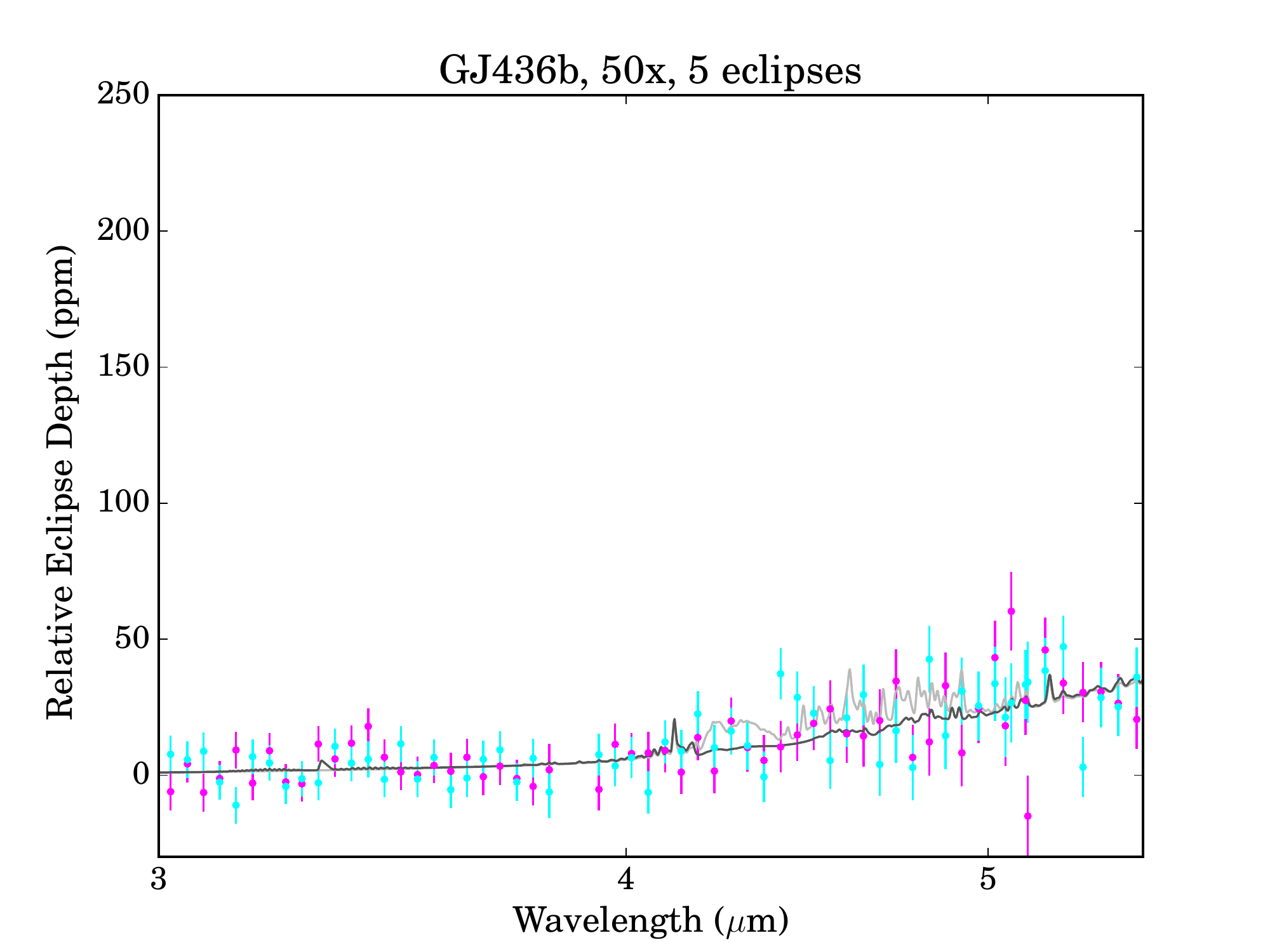}
\includegraphics[width=0.45\linewidth]{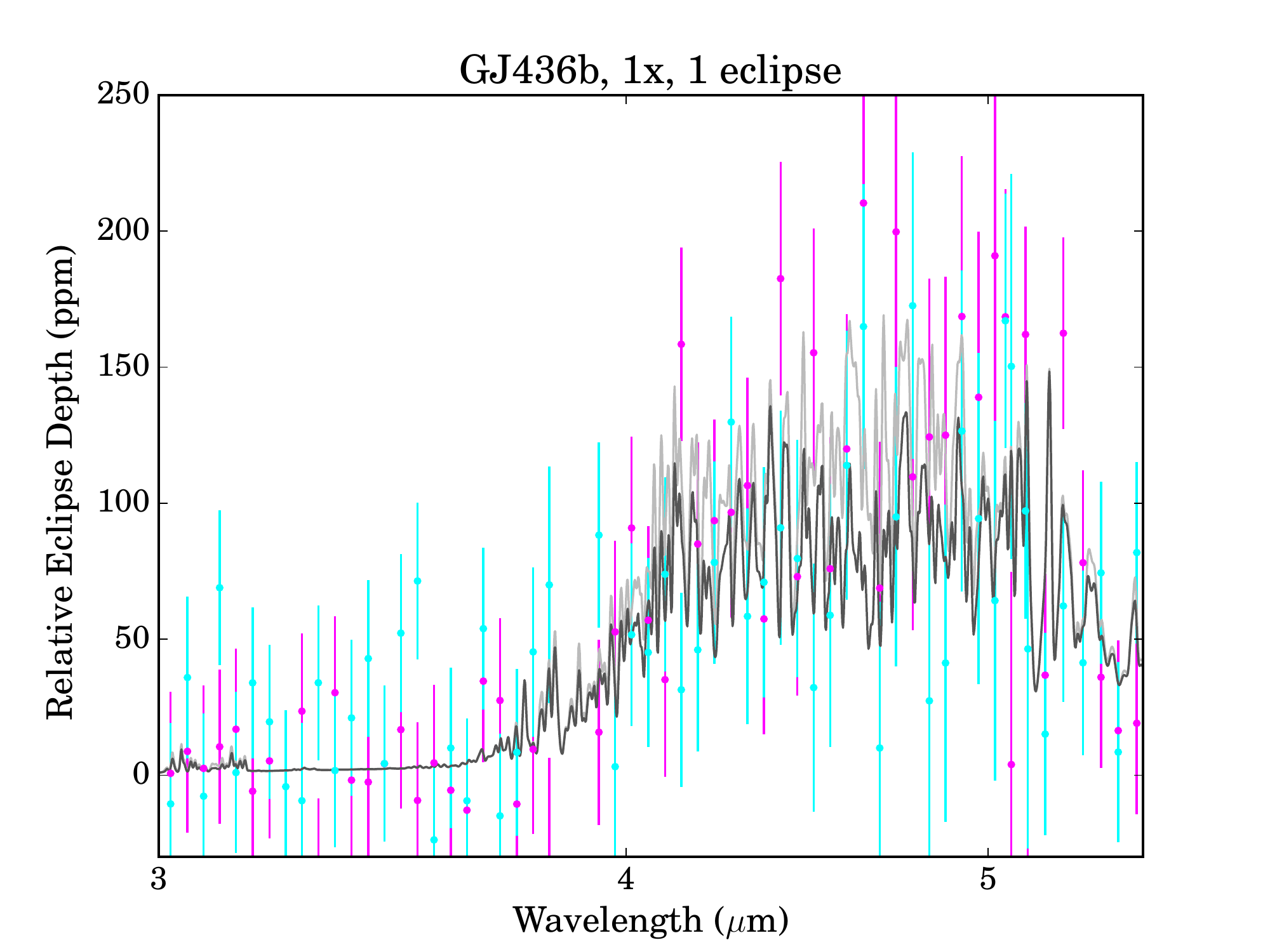}
\includegraphics[width=0.45\linewidth]{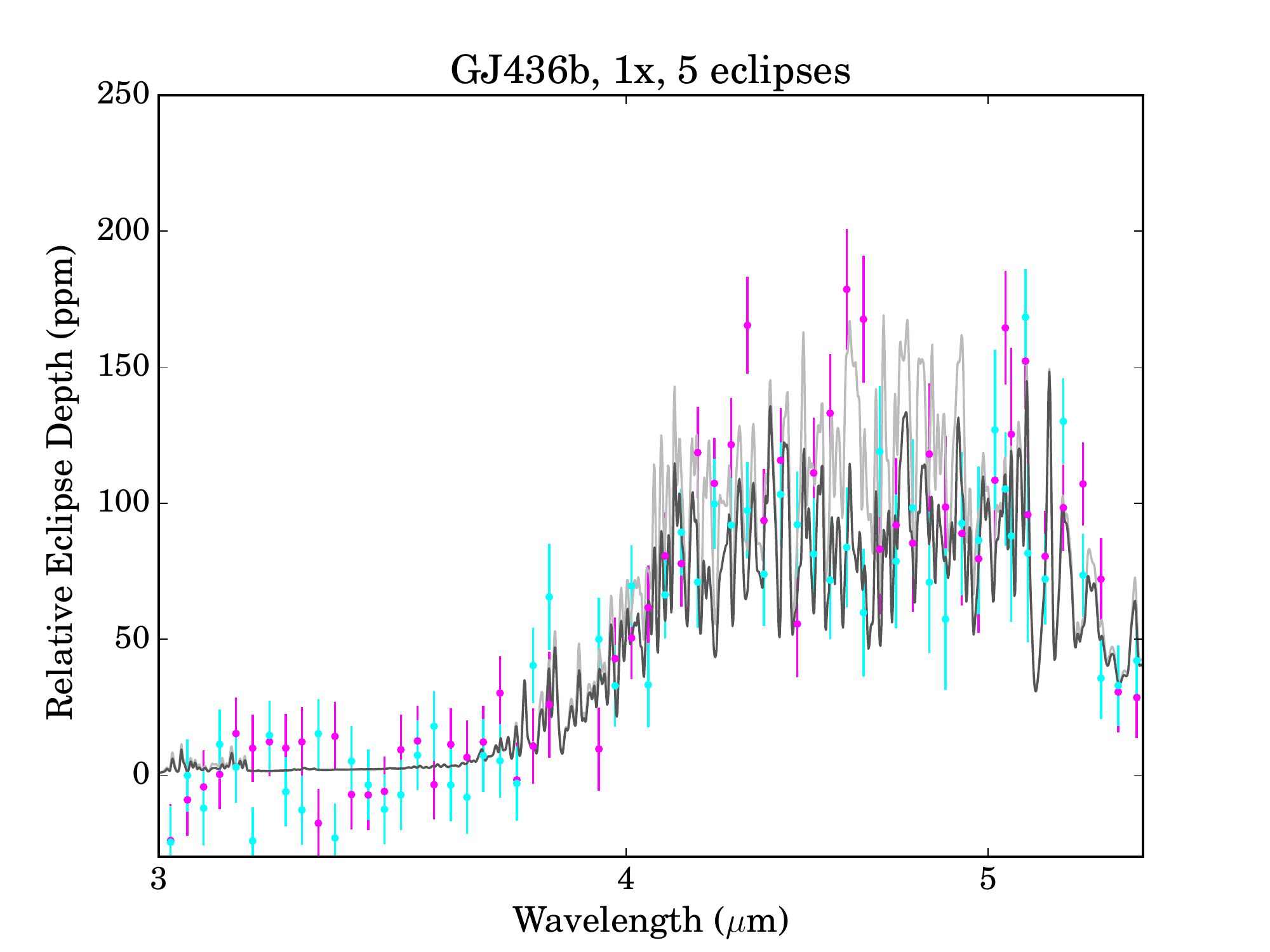}
\caption{Zoomed region of JWST simulated spectra for the GJ 436b-like planet to highlight the observable differences seen from 4 to 5 $\mu$m at R=50. Equilibrium chemistry points in magenta, high resolution model in light grey, and disequilibrium points in cyan, high resolution spectra in darker grey. Modelled with the \textit{row 1}: Kr14 metallicity (50$\times$), and \textit{row 2}: solar metallicity, for \textit{left column}: 1 eclipse event and \textit{right column}: 5 eclipse events.}
\label{fig:JWST_speczoom_GJ436b}
\end{figure*}

As time observing with JWST is at an ultimate premium, we examine how a few events can be used to make this distinction in our study for the wavelength region from 4 to 5 $\mu$m. We also choose to demonstrate the observability below native resolution (R=50) for emphasis. In Figures \ref{fig:JWST_speczoom_WASP80b} and \ref{fig:JWST_speczoom_GJ436b} the spectra calculated assume 1 and 5 eclipse events for both metallicities of the WASP-80b-like  and GJ 436b-like planets, respectively.  As expected, the more events, the clearer the separation between equiliabrium and disequilibrium in this region. For the WASP-80b-like planet (Figure \ref{fig:JWST_speczoom_WASP80b}, we can see the divergence after only 1 eclipse event. This divergence is clearer for solar metallicity case than for 5$\times$ solar metallicity case. Thus, with 5 eclipse events, the difference between equilibrium and disequilibrium spectra is clear in this wavelength region.  For the GJ 436b-like planet (Figure \ref{fig:JWST_speczoom_GJ436b}), we can only see this clear separation at solar metallicity and with 5 eclipses. There is too much noise to make the distinction between equilibrium and disequilibrium for 1 eclipse event at solar metallicity, and at high metallicity the distinction cannot be made at all. We choose to include this null result to show the effect of high metallicity.  Here it is important to discuss the assumed metallicity for our GJ 436b-like planet.  We assume a metallicity of 50$\times$ solar following  \citet{Kreidberg2014}, rather than a higher metallicity as reported in \citet{Moses2013GJ436b}. \citet{Moses2013GJ436b} singularly studies this planet and reports metallicity values of $\sim$230 to 2000 times solar. At these higher metallicities, we would also expect for the differences to be also undistinguishable given the rest of our model inputs. 

\begin{figure*}
\centering
\includegraphics[width=0.99\linewidth, clip]{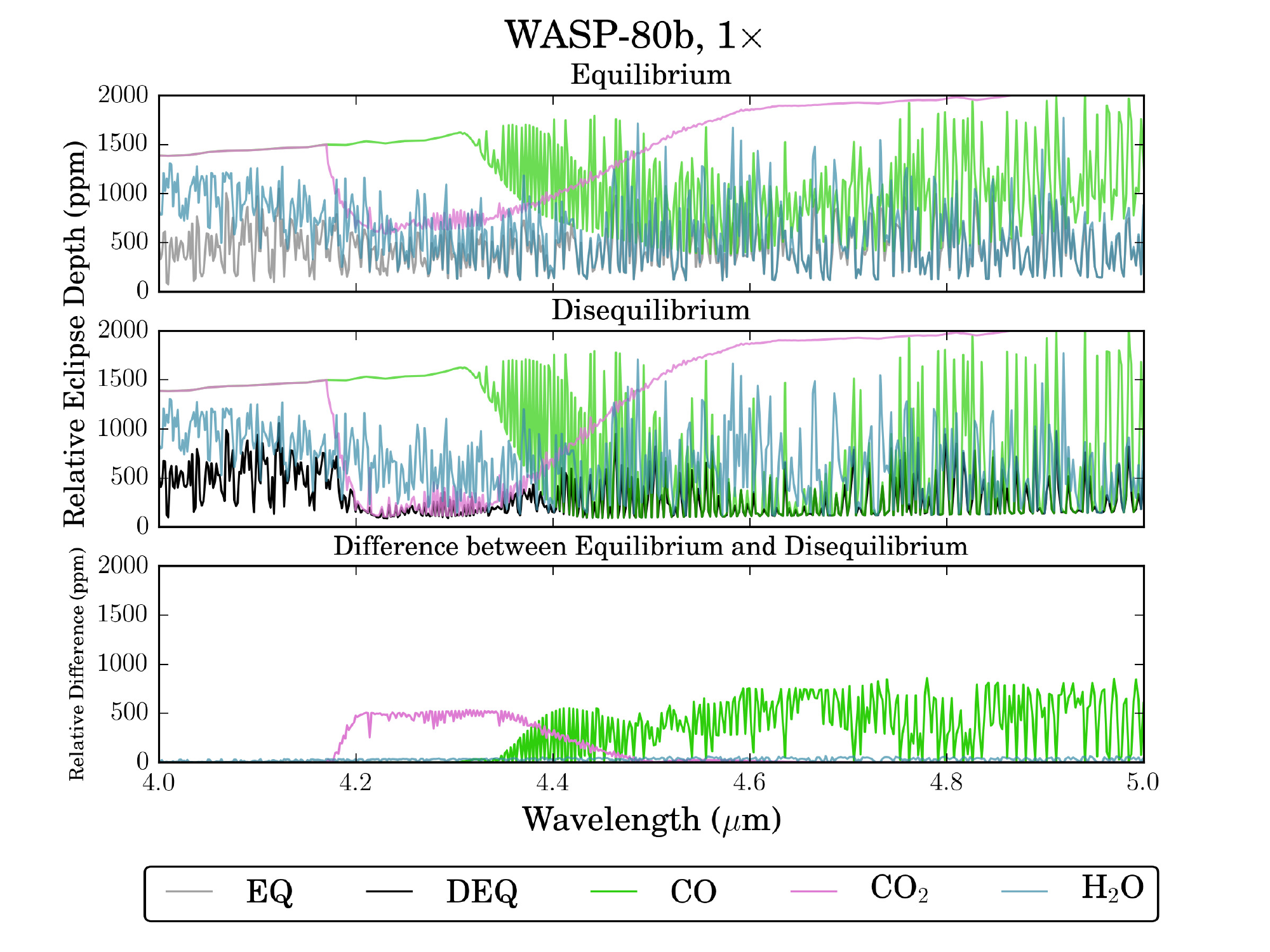}
\caption{Individual spectral contributions of CO (green), CO$_{2}$ (pink), and H$_{2}$O (blue) from disequilibrium chemistry (\textit{panel 1}) and equilibrium chemistry (\textit{panel 2}) overlaid on the overall spectra of disequilibrium (dark grey) and equilibrium (light grey) chemistry cases for the WASP-80b-like planet at solar metallicity. \textit{Panel 3} plots the differences between the disequilibrium and equilibrium cases for the aforementioned molecules.}
\label{fig:individ_WASP80b_1x}
\end{figure*}

\begin{figure*}
\centering
\includegraphics[width=0.99\linewidth, clip]{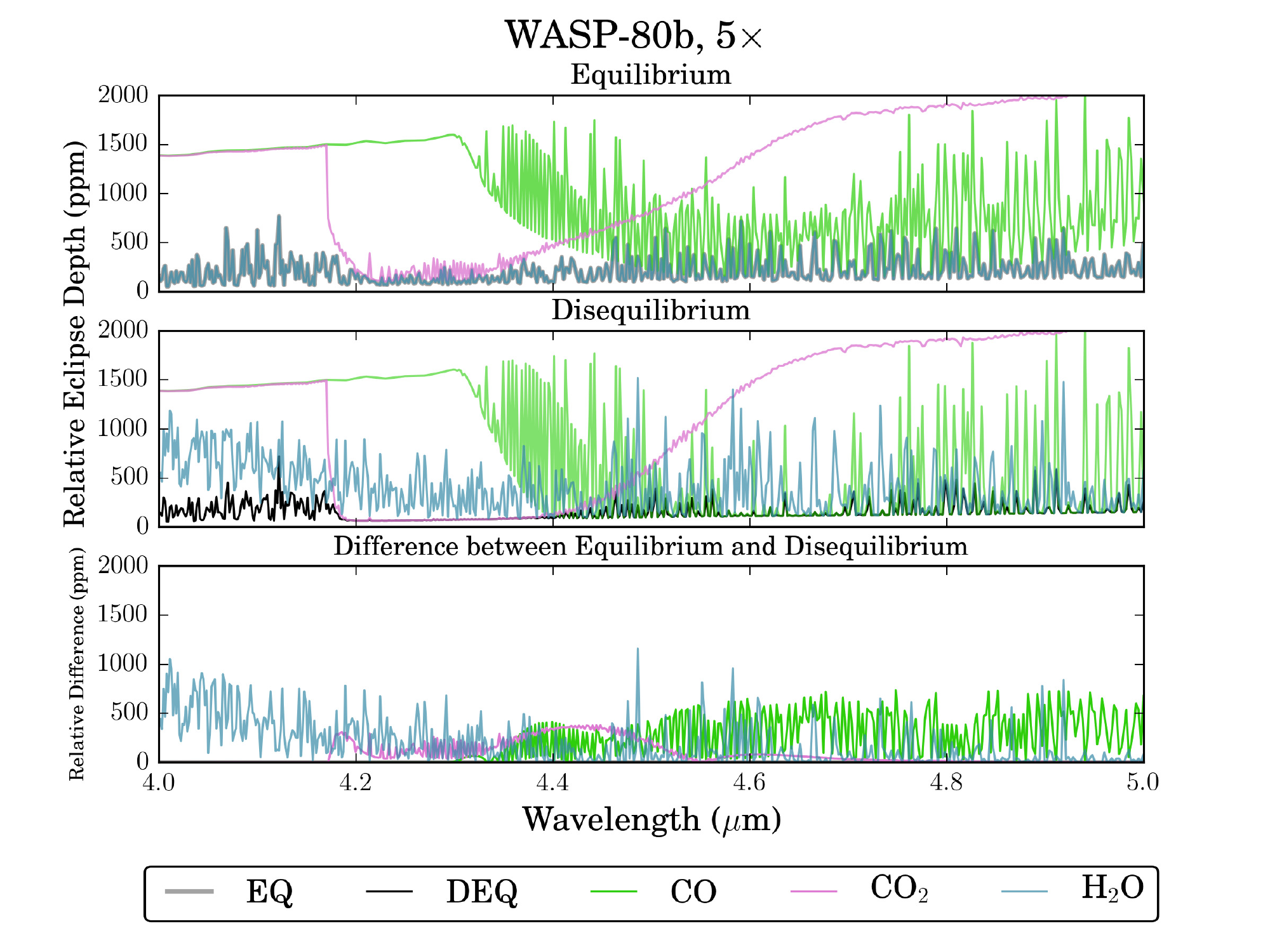}
\caption{Individual spectral contributions of CO (green), CO$_{2}$ (pink), and H$_{2}$O (blue) from disequilibrium chemistry (\textit{panel 1}) and equilibrium chemistry (\textit{panel 2}) overlaid on the overall spectra of disequilibrium (dark grey) and equilibrium (light grey) chemistry cases for the WASP-80b-like planet at 5 $\times$ metallicity. \textit{Panel 3} plots the differences between the disequilibrium and equilibrium cases for the aforementioned molecules.}
\label{fig:individ_WASP80b_5x}
\end{figure*}

\begin{figure*}
\centering
\includegraphics[width=0.99\linewidth, clip]{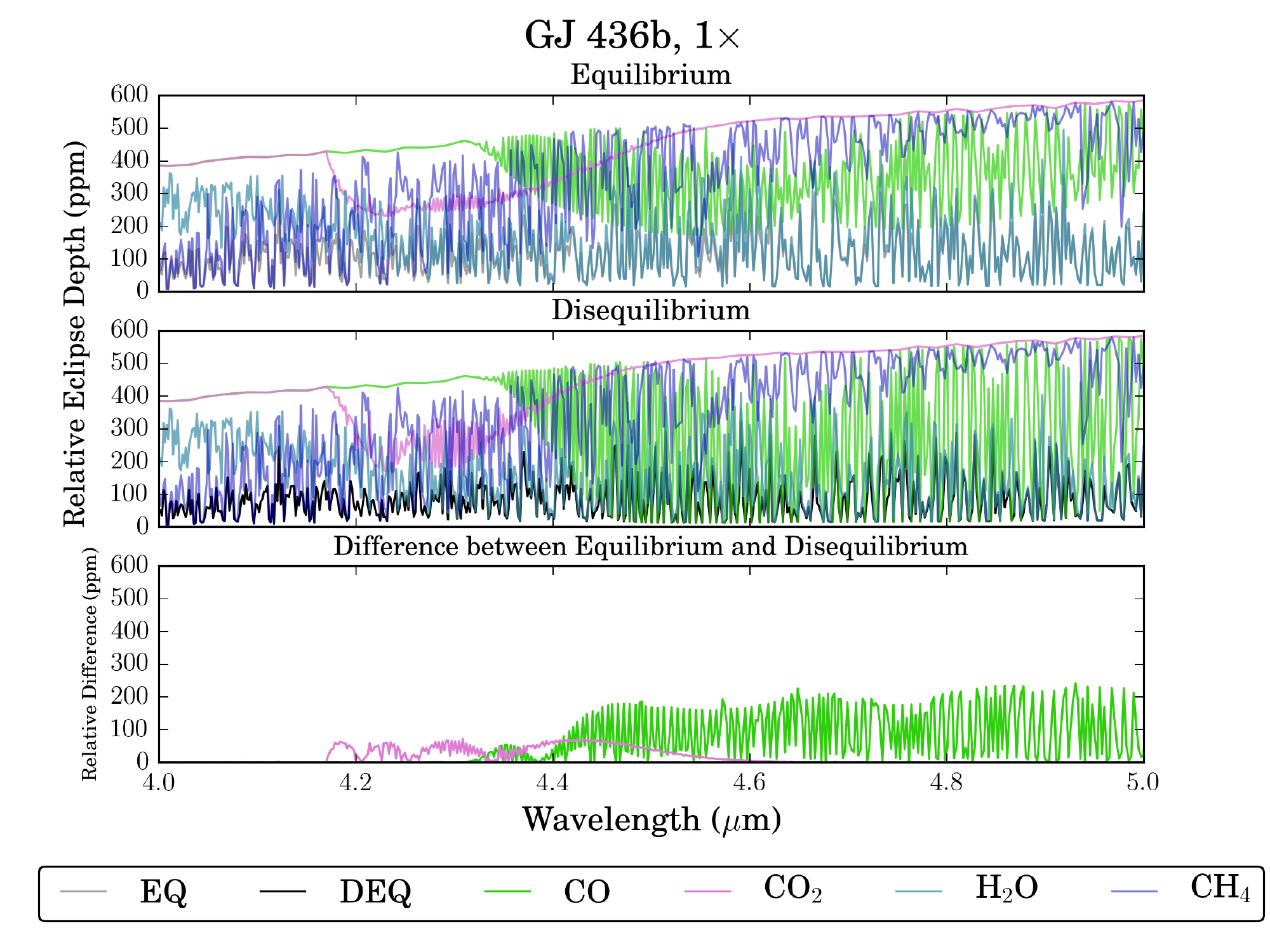}
\caption{Individual spectral contributions of CO (green), CO$_{2}$ (pink), and H$_{2}$O (blue) from disequilibrium chemistry (\textit{panel 1}) and equilibrium chemistry (\textit{panel 2}) overlaid on the overall spectra of disequilibrium (dark grey) and equilibrium (light grey) chemistry cases for the GJ 436b-like planet at solar metallicity. \textit{Panel 3} plots the differences between the disequilibrium and equilibrium cases for the aforementioned molecules.}
\label{fig:individ_GJ436b_1x}
\end{figure*}

We investigate the chemical sources of the difference seen from 4 to 5 $\mu$m by analyzing each species' contribution to the overall spectra. We discuss the contributions of individual species seen in Figures \ref{fig:individ_WASP80b_1x}, \ref{fig:individ_WASP80b_5x}, and \ref{fig:individ_GJ436b_1x}. Please note that the y-axis scale is kept consistent for each planet regardless of metallicity case. 

For the WASP-80b-like planet at solar metallicity, the spectral contributions from CO, CO$_{2}$, and H$_{2}$O predominantly comprise the overall disequilibrium and equilibrium spectra. For the disequilibrium chemistry case, the region from 4.2 to 4.4 $\mu$m is from CO$_{2}$, and the region from 4.4 to 5.0 $\mu$m is from CO. For the equilibrium case, the region from 4.3 to 5.0 $\mu$m is from H$_{2}$O. In Figure \ref{fig:individ_WASP80b_1x}, the overall difference between equilibrium and disequilibrium for the individual species does not represent a complete census on which species are responsible for the difference seen in Figure \ref{fig:JWST_spec} between 4.0 and 5.0 $\mu$m. Instead, the responsibility of this difference is from the relationship between these three species--- CO, CO$_{2}$, and H$_{2}$O. We can see that the difference between equilibrium and disequilibrium H$_{2}$O is small, but because water is so strongly absorbing, in the equilibrium case, its contribution masks the contributions of CO and CO$_{2}$. Thus, the difference we see between the overall equilibrium and disequilibrium spectra is from H$_{2}$O and disequilibrium CO and CO$_{2}$. In Figure \ref{fig:chem_majspec}, equilibrium CO and CO$_{2}$ are in are in less abundant than in disequilibrium--the largest divergence from equilibrium is around 10 mbar. As water abundance changes only slightly in the equilibrium and disequilibrium chemistry cases, the point at which water becomes optically thick is important as it could mask or reveal the divergence from equilibrium for CO and CO$_{2}$. 

Similar to the solar metallicity case for the WASP-80b-like planet, the 5$\times$ metallicity planet (Figure \ref{fig:individ_WASP80b_5x}) the contribution of disequilibrium CO$_{2}$ from 4.2 to 4.4 $\mu$m absorbs stronger than disequilibrium H$_{2}$O. However, for this metallicity case both disequilibrium CO and H$_{2}$O are responsible for the 4.4. to 5.0 $\mu$m region of the disequilibrium spectrum region. For the overall equilibrium spectrum, CO$_{2}$ and CO graze the overall spectrum, but this region from 4.0 to 5.0 $\mu$m, the equilibrium spectrum  is dominated by the contribution from equilibrium H$_{2}$O. In the difference plot in Figure \ref{fig:individ_WASP80b_5x} there is a larger difference between equilibrium and disequilibrium H$_{2}$O than at solar metallicity. Thus, the overall difference we see in Figure \ref{fig:JWST_spec} between 4.0 and 5.0 $\mu$m is between disequilibrium CO$_{2}$, CO, and H$_{2}$O and equilibrium H$_{2}$O. As this is a higher metallicity case, there is more H$_{2}$O in the system than at solar metallicity, making the difference between equilibrium and disequilibrium more spectrally relevant.

We examine the individual species' contributions for only the solar metallicity case of our GJ 436b-like planet as the difference between equilibrium and disequilibrium for 50$\times$ solar metallicity case is not observable. For this planet we have included the spectral contribution from CH$_{4}$ as it is relevant for this case but not for both metallicity cases of the WASP-80b-like planet. In Figure \ref{fig:individ_GJ436b_1x} from 4 to 4.4 $\mu$m the disequilibrium spectrum is dominated by disequilibrium CH$_{4}$, obscuring the contributions from H$_{2}$O and CO$_{2}$. For both disequilibrium and equilibrium CO$_{2}$ is not expressed in the overall spectra. From 4.4 to 5.0 $\mu$m, disequilibrium CO dominates the overall disequilibrium spectrum. For the equilibrium spectrum, again CH$_{4}$ dominates the same region from 4.0 to 4.4 $\mu$m, and equilibrium H${2}$O dominates the rest of the equilibrium spectrum from 4.4 to 5.0 $\mu$m. In the difference plot in Figure \ref{fig:individ_GJ436b_1x} there are large differences for CO but not for H$_{2}$O or CH$_{4}$. Thus, the difference between equilibrium and disequilibrium from 4.0 to 5.0 $\mu$m in Figure \ref{fig:JWST_spec}  is between disequilibrium CO and equilibrium H$_{2}$O. The increased abundance of CO in disequilibrium impacts the observed overall spectra. From Figure \ref{fig:chem_majspec}, we see large divergences from equilibrium for both CO and CO$_{2}$ centered at 10mbar, however because of the strong absorption from both CH$_{4}$ and H$_{2}$O, the spectral contribution from CO$_{2}$ is masked. Thus, similarly to the solar metallicity case of the WASP-80b-like planet, the point at which H$_{2}$O becomes optically thick will determine if the contribution from CO is masked or revealed in the overall observed spectra.

\subsection{Further Discussion and Future Work}

\citet{Burrows2014} points out the ``primitive''  state of exoplanet data, and therefore how pinning down metallicity is difficult at this point. \citet{Burrows2014} explains that until higher resolution data across a large wavelength domain is available,  only then will we be able to start making better conclusions about metallicity. Due to the increase of resolution with JWST, it is important to take disequilibrium chemistry into account when determining metallicity. In particular, in this study we see that the abundances of CO, CO$_{2}$, and H$_{2}$O are sensitive to disequilibrium chemistry and that their contributions are create an observable difference between 4.0 and 5.0 $\mu$m. \citet{MadhuSeager2011} points out that for their study of GJ 436b, chemical abundances cannot be fully explained using equilibrium chemistry, and therefore the metallicity cannot be determined using these equilibrium abundances. By taking disequilibrium chemistry (including photochemistry), we can better determine the metallicity of giant gaseous exoplanets. 

We choose to model only two metallicities per planet-- solar and Kr14 metallicity (as mentioned in \S\ref{sec:method}). \citet{Kreidberg2014} suggests that atmospheric metallicity of giant gas planets in the Solar System might be significantly higher than the stellar metallicity. The relationship derived in \citet{Kreidberg2014} uses CH$_{4}$ to derive Solar System metallicities and H$_{2}$O for exoplanet metallicity. However, \citet{GnG2014} points out that the molecular abundances and therefore bulk metallicity for Solar Systems objects are notoriously hard to pin down. \citet{GnG2014} also points out the vast progression over the mere last ten years in constraining the Sun's elemental composition; and the non-linear divergence from solar metallicity in our own gas planets reporting that in carbon alone that it enriched 4 fold in Jupiter, 10 fold in Saturn, and 90 fold in Neptune. Additionally, the Galileo probe found on Jupiter that carbon, nitrogen and sulfur's elemental abundances were supersolar between 2.5 to 4.5 times \citep{Wong2004}. As for oxygen, its elemental abundance is harder to constrain as water is so much out of equilibrium on Jupiter (and the probe fell into a hot presumably dry spot and did not finish it's measurements). For exoplanet analysis, it has been a popular assumption to use solar metallicity \citep{CooperShowman2006, Fortney2006, Fortney2007, Fortney2008, Venot2012, Moses2011, Koskinen2013, ZahnleMarley2014, Wakeford2015, Heng2016, Kataria2016, Barstow2017, Gao2017}, or a linear scaling of solar metallicity  \citep{Freedman2014, Kataria2015, Charnay2015, Wakeford2017}, and hence why we chose to model it as one of our metallicity cases. The bulk composition of a planet's atmosphere plays a major role in the chemistry at work in the planet's atmosphere. The bulk composition of exoplanet atmospheres will depend on its formation history \citep{Oberg2011, Madhu2014, Wakeford2017}. \citet{Wakeford2017} clearly demonstrates this with the analysis of HAT-P-26b which departs from the Kr14 relationship. Thus, we recognize the limited metallicity space we have explored but have accomplished how disequilibrium chemistry should be taken into account when determining metallicity and how JWST may provide insight into this determination. By better understanding our own giant planets's non-linear relationship with solar elemental abundance, it may help us to understand our own system's formation. From there, we may be able to better diagnose the bulk composition of exoplanets and understand their formation. (The \textit{Juno} mission is currently attempting to discern the H$_{2}$O abundance of Jupiter using gravitational and radar diagnostics. )  Retrieval methods may benefit from exploring and/or iterating over a diverse set of scenarios for a planet's bulk elemental abundances. We leave exploration of more metallicities including non-linear scaling of solar elemental abundance for future studies that seeks to match observations. 

It is also important here to remark on the reliance of the abundances of CO, CO$_{2}$, or H$_{2}$O on the TP profile. As mentioned in Section \S\ref{sec:method}, our TP profile is not calculated self-consistently. In order to isolate the differences between equilibrium and disequilibrium chemistry, we kept the TP profiles identical whereas consistently-calculated profiles differ based on the whether the chemistry is specified as in or out of equilibrium. Given the work of \citet{Drummond2016} which models TP profiles self-consistently in and out of equilibrium, the resultant TP profiles of HD 189733b would probably lie in-between equilibrium and non-equilibrium (also called disequilibrium) chemistry profiles. Thus, as the profiles presented in this study are calculated analytically with no radiative feedback, our disequilibrium results would be affected. With self-consistently calculated TP profiles, our disequilibrium results would also probably lie somewhere in between our current equilibrium and disequilibrium results. 
 
We recognize the limitations of a 1D study for accurate comparison with observations, but we utilize 1D here for rapid exploration of parameter space. There has been limited work employing chemical kinetics in 3D. \citet{CooperShowman2006} study CO, CH$_{4}$, and H$_{2}$O in 3D with a limited chemical network, incorporating the chemical kinetic rate of only the rate-limiting reactions into the relaxation timescale.  In between 1D and 3D is the work of \citet{Agundez2014} which studies HD 189733b and HD 209458b in `pseudo 2D.' \citet{Agundez2014} uses a robust chemical scheme and simplified dynamics to study the interplay between chemistry and dynamics. Again, we leave modelling chemical kinetics in 3D for future work that seeks to match observations. 

Finally, the chemistry models we use in this study, although considered state-of-art, contain only gaseous species for carbon, nitrogen, and oxygen chemistry. As pointed out in \citet{Zahnle2009, Zahnle2016} the inclusion of sulfur-containing species is important however the reaction rates for these species at high temperature remain poorly constrained. Additionally, \citet{Fortney2016Lab} points that we also lack the optical data to interpret these sulfur containing gases. It is crucial to point out that our study does not include haze or clouds and the results presented here could be seriously skewed without their inclusion \citep{Morley2012, Morley2015, Wakeford2015, Sing2016, Parmentier2016, Line2016, Wakeford2017}. Future, more realistic studies will include condensed phase chemistry.

\section{Conclusions}
\label{sec:discussion}
 
In our study we seek not to match observations, but explore a limited parameter space to demonstrate the observability of disequilibrium chemistry with JWST in 1D. We provide a 1D study that for the first time links equilibrium versus disequilibrium chemistry to simulated JWST spectra.  We find a sweet spot in radius ($>$ 0.952R$_{Jup}$), metallicity (between 1$\times$ and 5$\times$ solar), and temperature ($<  \sim$800 K) for observing differences due to disequilibrium chemistry from 4 to 5 $\mu$m with the NIRSpec G395M. We find that the spectral signature of an atmosphere in chemical disequilibrium can be distinguished from an atmosphere in chemical equilibrium for a planet like warm-Jupiter WASP-80b for the two input metallicity cases; but for the slightly cooler sub-Neptune GJ 436b we find that we can only partially detect the contribution from disequilibrium chemistry in the case of solar metallicity but not for the 50$\times$ solar metallicity case. We report that based on a select set of input parameters that CO, CO$_{2}$, and H$_{2}$O are responsible for differences in our WASP-80b-like and GJ 436b-like planet cases. Thus, as the signatures from CO and CO$_{2}$ are visible in the observed disequilibrium spectra, but not in equilibrium spectra, disequilibrium chemistry should be considered when determining the metallicity of a planet. We can not observe disequilibrium chemistry in the HD 189733b-like planet given our inputs because it is too hot. At such high temperatures, thermochemical equilibrium dominates over disequilibrium processes.

\section*{Acknowledgments}

SDB would like to thank Ned Molter and Conor Nixon for meaningful conversations about hydrostatic equilibrium, and Aarynn Carter for meaningful conversations about JWST.
SDB thanks NASA GSFC and UMBC for support of this work, and the University of Exeter for support through a Ph.D. studentship. Finally, SDB would like to thank the referee whose comments greatly improved the manuscript.

\end{document}